\documentclass[10pt,oneside,openright,a4paper,reqno]{amsart}
\usepackage{times}
\usepackage{hyperref}
\usepackage[foot]{amsaddr}
\usepackage{amsmath}
\usepackage{amsfonts, mathrsfs}
\usepackage{amssymb}
\usepackage{paralist}
\usepackage{algorithmic,algorithm}
\usepackage{float}
\usepackage{graphicx}
\usepackage{adjustbox}
\usepackage{enumerate}

\newtheorem{anytheorem}{Theorem}[section]

\newtheorem{theorem}[anytheorem]{Theorem}
\theoremstyle{definition}

\newtheorem{example}[anytheorem]{Example}

\DeclareMathOperator*{\argmin}{arg\,min}

\title{Unbiased deep solvers for linear parametric PDEs}
\author{Marc Sabate Vidales$^{1,2}$}
\email{M.Sabate-Vidales@sms.ed.ac.uk}
\author{David \v{S}i\v{s}ka$^{2,3}$}
\email{D.Siska@ed.ac.uk}
\author{Lukasz Szpruch$^{1,2}$}
\address{$^1$\href{https://www.turing.ac.uk}{The Alan Turing Institute}\\ 
    $^2$\href{https://www.maths.ed.ac.uk}{University of Edinburgh School of Mathematics}\\$^3$\href{https://vega.xyz}{Vega}}
\email{L.Szpruch@ed.ac.uk}

\date{\today}

\begin{document}

\begin{abstract}
We develop several deep learning algorithms for approximating families of parametric PDE solutions.
The proposed algorithms approximate solutions together with their gradients, which in the context of mathematical finance means that the derivative prices
and hedging strategies are computed simulatenously. 
Having approximated the gradient of the solution one can combine it with a Monte-Carlo simulation to remove the bias in the deep network approximation of the PDE solution (derivative price). 
This is achieved by leveraging the Martingale Representation Theorem and combining the Monte Carlo simulation with the neural network.
The resulting algorithm is robust with respect to quality of the neural network approximation and consequently can be used as a black-box in case only limited a priori information about the underlying problem is available. We believe this is important as neural network based algorithms often require fair amount of tuning to produce satisfactory results.
The methods are empirically shown to work for high-dimensional problems (e.g. 100 dimensions).
We provide diagnostics that shed light on appropriate network architectures. 
\end{abstract}




\maketitle
\setcounter{tocdepth}{1}

\section{Introduction}

Numerical algorithms that solve PDEs suffer from the so-called ``curse of dimensionality", making it impractical to apply known discretisation algorithms such as finite differences schemes to solve high-dimensional PDEs. However, it has been recently shown that deep neural networks trained with stochastic gradient descent can overcome the curse of dimensionality~\cite{beck2020overcoming, Berner_2020}, making them a popular choice to solve this computational challenge in the last few years.

In this work, we focus on the problem of numerically solving parametric linear PDEs arising from European option pricing in high-dimensions. 
Let $B\subseteq \mathbb R^p, p\geq 1$ be a parameter space (for instance, in the Black--Scholes equation with fixed rate, $B$ is the domain of the volatility parameter). 
Consider $v = v(t,x; \beta)$ satisfying
\begin{equation}\label{eq pde intro}
\begin{split}
& \bigg[\partial_t v + b \nabla_x v + \frac{1}{2}\text{tr}\left[\nabla_x^2 v \sigma^*\sigma\right] - cv\bigg](t,x;\beta)  = 0\,, \\
& v(T,x;\beta)  = g(x;\beta)\,,\,\,\, t \in [0,T]\,,\,\, x\in \mathbb R^d\,,\,\,\beta \in B\,.
\end{split}
\end{equation} 
Here $t\in [0,T]$, $x\in\mathbb R^d$ and $\beta\in B$ and $b,\sigma, c$ and $g$ are functions of $(t,x;\beta)$ which specify the problem. The Feynman--Kac theorem provides a probabilistic representation for $v$ so that Monte Carlo methods can be used for its unbiased approximation in one single point $(t,x;\beta)$.  What we propose in this work is a method for harnessing the power of deep learning algorithms to numerically solve~\eqref{eq pde intro} in a way that is robust even in edge cases when the output of the neural network is not of the expected quality, by combining them with Monte Carlo algorithms.

From the results in this article we observe that neural networks provide an
efficient computational device for high dimensional problems.
However, we observed that these algorithms are sensitive to the network
architecture, parameters and distribution of training data. 
A fair amount of tuning is required to obtain good results. 
Based on this we believe that there is great potential in combining artificial
neural networks with already developed and well understood probabilistic 
computational methods, in particular the control variate method for using potentially imperfect neural network approximations for finding unbiased solutions to a given problem, see Algorithm~\ref{alg unbiased parametric pde solver}.

\subsection{Main contributions}
We propose three classes of learning algorithms for simultaneously 
finding solutions and gradients to parametric families of PDEs.
 
\begin{enumerate}[i)]
\item {\em Projection solver}: See Algorithm~\ref{alg PDE cond expec}. We leverage Feynman--Kac representation together with with the fact 
that conditional expectation can be viewed as an $L^2$-projection operator. 
The gradient can be obtained by automatic differentiation of already obtained approximation of the PDE solution.
\item {\em Martingale representation solver}:
See Algorithm~\ref{alg PDE prob repr iterative training}. 
This algorithm was inspired by Cvitanic et. al.~\cite{cvitanic2005steepest} and Weinan et. al, Han et. al.~\cite{weinan2017deep,han2017solving} and is referred to as deep BSDE solver. 
Our algorithm differs from~\cite{weinan2017deep} in that we approximate solution and its gradient at all the time-steps and across the entire space and parameter domains rather than only one space-time point. 
Furthermore we propose to approximate the solution-map and its gradient by separate networks. 
\item {\em Martingale control variates solver}: Algorithms~\ref{alg empirical risk minimisation} and~\ref{alg empirical corr maximisation}. 
Here we exploit the fact that martingale representation induces control variate that can produce zero variance estimator. Obviously, such control variate is not implementable but provides a basis for a novel learning algorithm for the PDE solution. 
\end{enumerate}

For each of these classes of algorithms we develop and test different implementation strategies. Indeed, one can either take one (large)
network to approximate the entire family of solutions of~\eqref{eq pde intro} or take
a number of (smaller) networks, where each of them approximates 
the solution at a time point in a grid.
The former has the advantage that one can take arbitrarily fine time discretisation 
without increasing the overall network size.
The advantage of the latter is that each learning task is simpler due 
to each network being smaller. 
One can further leverage the smoothness of the solution in time and 
learn the weights iteratively by initialising the network parameters to be those of the previous time step. 
We test both approaches numerically. At a high level all the algorithms work in path-dependent (non-Markovian) setting but there the challenge is an efficient method for encoding information in each path. 
This problem is solved in companion paper~\cite{sabatevidales2020solving}.

\begin{figure}
\includegraphics[width=0.5\linewidth]{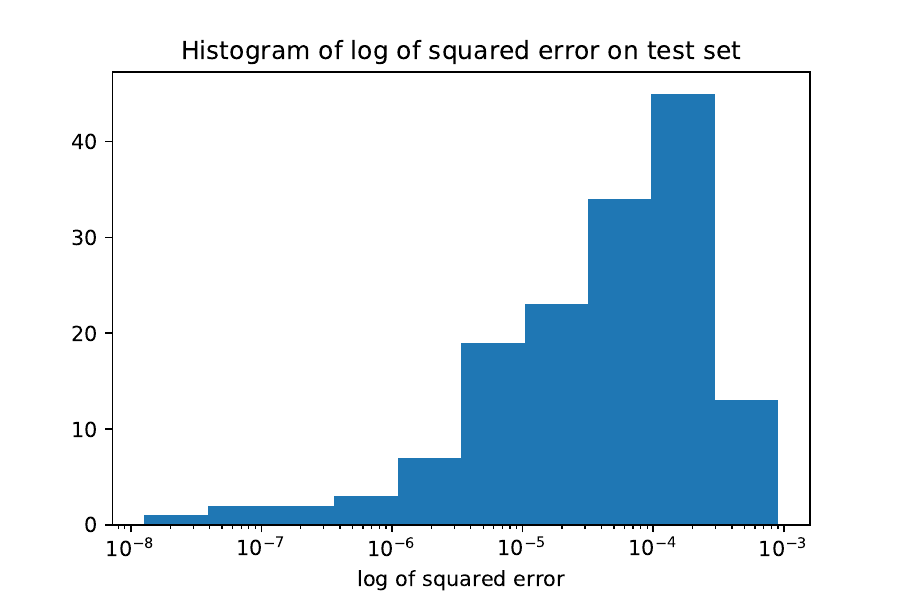}
\caption{Histogram of mean-square-error of solution to the PDE on the test data set.}
\label{fig hist MSE test set}
\end{figure}

To summarise the key contribution of this work are:
\begin{enumerate}[i)]
\item We derive and implement three classes of learning algorithms for approximation of parametric PDE solution map and its gradient. 
\item We propose a novel iterative training algorithm that exploits regularity of the function we seek to approximate and allows using neural networks with smaller number of parameters.
\item \label{item black box} The proposed algorithms are truly black-box in that quality of the network approximation only impacts the computation benefit of the approach and does not introduce approximation bias. 
This is achieved by combining the network approximation with Monte Carlo as stated in Algorithm~\ref{alg unbiased parametric pde solver}.
\item Code for the numerical experiments presented in this paper is being made available on GitHub: \texttt{\href{https://github.com/msabvid/Deep-PDE-Solvers}{https://github.com/msabvid/Deep-PDE-Solvers}}.
\end{enumerate}
We stress the importance of point~\ref{item black box}) above by directing reader's attention to Figure~\ref{fig hist MSE test set}, where we test generalisation error of trained neural network for the 5 dimensional family of PDEs corresponding to pricing a basket option under the Black--Scholes model. We refer reader to Example~\ref{ex basket random sigma} for details. 
We see that while the average error over test set is of order $\approx10^{-5}$, the errors for a given input varies significantly. 
Indeed, it has been observed in deep learning community that for high dimensional problems one can find input data such that trained neural network that appears to generalise well (i.e achieves small errors on the out of training data) produces poor results \cite{goodfellow2014explaining}.

\subsection{Literature review}

Deep neural networks trained with stochastic gradient descent proved to be extremely successful in number of applications such as computer vision, natural language processing, generative models or reinforcement learning~\cite{lecun2015deep}. 
The application to PDE solvers is relatively new and has been pioneered by Weinan et. al, Han et. al.~\cite{weinan2017deep,han2017solving,sirignano2017dgm}. 
See also Cvitanic et. al.~\cite{cvitanic2005steepest} for the ideas of solving PDEs with gradient methods and  for direct PDE approximation algorithm. PDEs provide an excellent test bed for neural networks approximation because a) there exists alternative solvers e.g Monte Carlo b) we have well developed theory for PDEs, and that knowledge can be used to tune algorithms. 
This is contrast to mainstream neural networks approximations in text or images classification. 

Apart from growing body of empirical results in literature on ``Deep PDEs solvers'', \cite{chan2019machine,hure2019some,beck2018solving,jacquier2019deep,henry2017deep} there has been also some important theoretical contributions. It has been proved that deep artificial neural networks approximate solutions to parabolic PDEs to an arbitrary accuracy without suffering from the curse of dimensionality. 
The first mathematically rigorous proofs are given in~\cite{GrohsHornungJentzenvonWurstemberger2018}
and~\cite{jentzen2018proof}. The high level idea is to show that neural network approximation to the PDE can be established by building on Feynman-Kac approximation and Monte-Carlo approximation. 
By checking that Monte-Carlo simulations do not suffer from the curse of dimensionality  one can imply that the same is true for neural network approximation.
Furthermore, it has been recently demonstrated in \cite{hu2019mean,mei2018mean}  that noisy gradient descent algorithm used for training of neural networks of the form considered in \cite{GrohsHornungJentzenvonWurstemberger2018,jentzen2018proof} induces unique probability distribution function over the parameter space which minimises learning.
See \cite{du2018gradient,chizat2018global,rotskoff2018neural,sirignano2019mean,wang2021gradient,han2020convergence} for related ideas on convergence of gradient algorithms for overparametrised neural networks. 
This means that there are theoretical guarantees for the approximation of (parabolic) PDEs with neural networks trained by noisy gradient methods alleviating the curse of dimensionality.

An important application of deep PDE solvers is that one can in fact approximate a parametric family of solutions of a PDE.
To be more precise let $B \subseteq \mathbb R^p$, $p\geq 1$, be a parameter space.
In the context of finance these, for example, might be initial volatility, volatility of volatility, interest rate and mean reversion parameters. 
One can approximate the parametric family of functions $F(\cdot;\beta)_{\beta \in B}$ for an arbitrary range of parameters. 
This then allows for swift calibration of models to data (e.g options prices). 
This is particularly appealing for high dimensional problems when calibrating directly using noisy Monte-Carlo samples might be inefficient. 
This line of research gained some popularity recently and the idea has been tested numerically on various models and data sets~\cite{horvath2019deep,liu2019neural,bayer2018deep,stone2018calibrating,hernandez2016model,itkin2019deep,mcghee2018artificial}. 
There are some remarks that are in order. In the context of models calibration, while the training might be expensive one can do it offline, once and for good. One can also notice that training data could be used to produce a ``look-up table'' taking model parameters to prices. 
From this perspective the neural network, essentially, becomes an interpolator and a compression tool. 
Indeed the number of parameters of the network is much smaller than number of training data and therefore it is more efficient to store those. 
The final remark is that while there are other methods out there, such as Chebyshev functions, neural networks seem robust in high dimensions which make them our method of choice. 

\subsection{Notation} We denote by $\mathcal{DN}$ the set of all fully connected feedforward neural networks (see Appendix~\ref{sec dn}). We also use $\mathcal R[f]_\theta \in \mathcal{DN}$ with $\theta\in\mathbb R^{\kappa}$ to denote a neural network with weights $\theta$ approximating the function $f:\mathbb R^{d_0} \rightarrow \mathbb R^{d_1}$ for some $d_0, d_1\in\mathbb N$. 

\subsection{Outline}
This paper is organised as follows.
Section~\ref{sec martingale control var} provides theoretical underpinning 
for the derivation of all the algorithms we propose to solve~\eqref{eq pde intro}. More specifically in Section~\ref{sec unbiased pde solver} we combine the approximation of the gradient of the solution of the PDE resulting from the Deep Learning algorithms with Monte Carlo to obtain an unbiased approximation of the solution of the PDE.
In Section~\ref{sec deep pde solvers}, we describe the algorithms in detail.  


Finally in Section~\ref{sec examples} we provide numerical tests of the proposed algorithms.
We empirically test these methods on relevant examples including a 100 dimensional option pricing problems, see Examples~\ref{ex 100d average exchange} and~\ref{ex basket 100d}. 
We carefully measure the training cost and report the variance reduction 
achieved. 

Since we work in situation where the function approximated by neural network can be obtained via other methods (Monte-Carlo, PDE solution) we are 
able to test the how the expressiveness of fully connected artificial neural networks depends on the number of layers and neurons per layer.
See Section~\ref{sec experiments empirical net diag} for details.

\section{PDE Martingale control variate}
\label{sec martingale control var}

Control variate is one of the most powerful variance reduction techniques for Monte-Carlo simulation. 
While a good control variate can reduce the computational cost of Monte-Carlo computation by several orders of magnitude, 
it relies on judiciously chosen control variate functions that are problem specific. 
For example, when computing price of basket options a sound strategy is to choose control variates to be call options written on each of the stocks in the basket, since in many models these are priced by closed-form formulae.  
In this article, we are interested in black-box-type control variate approach by leveraging the Martingale Representation Theorem and neural networks. 
The idea of using Martingale Representation to obtain control variates goes back at least to~\cite{newton1994variance}. 
It has been further studied in combination with regression in~\cite{milstein2009solving} and~\cite{belomestny2018variance}. 

The bias in the approximation of the solution can be completely removed by employing control variates
where the deep network provides the control variate resulting in very high
variance reduction factor in the corresponding Monte Carlo simulation.

Let $(\Omega, \mathcal F, \mathbb P)$ be a probability space 
and consider an $\mathbb R^{d'}$-valued Wiener process $W=(W^j)_{j=1}^{d'} = ((W^j_t)_{t\geq 0})_{j=1}^{d'}$.
We will use $(\mathcal F^W_t)_{t\geq 0}$ to denote the filtration generated by $W$. 
Consider a $D\subseteq \mathbb R^d$-valued, continuous, stochastic process defined for the parameters $\beta\in B\subseteq\mathbb R^p$, $X^{\beta}=(X^{\beta,i})_{i=1}^d = ((X^{\beta,i}_t)_{t\geq 0})_{i=1}^d$ adapted to $(\mathcal F^W_t)_{t\geq 0}$ given as the solution to
\begin{equation}\label{eq sde}
dX_s^\beta = b(s,X_s^\beta; \beta)\,ds + \sigma(s,X_s; \beta)\,dW_s,\,\,\,s\in[t,T]\,,\,\,\, X_t^\beta = x \in \mathbb R^d.
\end{equation}
We will use $(\mathcal F_t^\beta)_{t\geq 0}$ to denote the filtration generated by $X^\beta$. 

Let $g : \mathbb R^d \to \mathbb R$ 
be a measurable function and we assume that there is a (stochastic) discount factor given by 
\[
D(t_1,t_2;\beta) := e^{-\int_{t_1}^{t_2} c(s,X_s^\beta;\beta)\,ds}
\]
for an appropriate function $c=c(t,x;\beta)$.
We will omit $\beta$ from the discount factor notation for brevity.
We now interpret $\mathbb P$ as some risk-neutral measure and 
so the $\mathbb P$-price of our contingent claim is 
\begin{equation}
\label{eq fk}
v(t,x;\beta) := \mathbb E \left[D(t,T) g(X_T^\beta) \bigg| X_t^\beta=x \right]\,.
\end{equation}

Say we have iid r.v.s $(X_T^{\beta,i})_{i=1}^{N}$ with the same distribution as $X_T^\beta$, where for each $i$, $X_t^{\beta,i}=x$.
Then the standard Monte-Carlo estimator is
\[
 v^N(t,x;\beta) := \frac{1}{N}\sum_{i=1}^{N}D^i(t,T) g(X_T^{\beta,i})\,.
\]
Convergence $v^N(t,x;\beta) \to v(t,x;\beta)$ in probability as $N\to\infty$ is granted by the Law of Large Numbers. 
Moreover the classical Central Limit Theorem tells that 
\[
\mathbb P\left(\! v(t,x;\beta) \! \in \! \left[ v^N(t,x;\beta) - z_{\alpha/2}\frac{\sigma}{\sqrt{N}}, 
v^N(t,x;\beta) + z_{\alpha/2}\frac{\sigma}{\sqrt{N}}  \right] \right) \rightarrow 1 - \alpha\,\,\text{as}\,\,N\rightarrow \infty\,,
\] 
where $\sigma:=\sqrt{ \mathbb{V}ar\left[D(t,T) g(X_T^\beta)\right] }$ and $z_{\alpha/2}$ is such that $1-\Phi(z_{\alpha/2}) = \alpha/2$ 
with $\Phi$ being distribution function (cumulative distribution function) of the standard normal distribution. 
To decrease the width of the confidence intervals one can increase $N$, but this also increases the computational cost. 
A better strategy is to reduce variance by finding an alternative Monte-Carlo estimator, say $\mathcal V^N(t,x; \beta)$, such that 
\begin{equation} \label{cv property}
\mathbb{E}[\mathcal V^N(t,x; \beta)] = v(t,x;\beta)\,\,\,\,\, \text{and} \,\,\,\,\, \mathbb{V}ar[\mathcal V^N(t,x;\beta)] < \mathbb{V}ar[v^N(t,x;\beta)], 
\end{equation}
and the cost of computing  $\mathcal V^N (t,x;\beta)$ is similar to $v^N (t,x;\beta)$.

In the remainder of the article we will devise and test several strategies, based on deep learning, to find a suitable approximation for $\mathcal V^N(t,x;\beta)$, by exploring the connection of the SDE~\eqref{eq sde} and its associated PDE. 

\subsection{PDE derivation of the control variate}\label{sec pde control variate}


It can be shown that under suitable assumptions on $b$, $\sigma$, $c$ and $g$, and fixed $\beta\in B$ that $v\in C^{1,2}([0,T]\times D)$.
See e.g.~\cite{krylov1999kolmogorov}. 
Let $a:=\frac12\sigma \sigma^*$. 
Then, from Feynman--Kac formula (see e.g. Th. 8.2.1 in ~\cite{oksendal2003stochastic}), we get
\begin{equation}\label{eq pde}
\left\{
\begin{split}
\left[\partial_t v + \text{tr}(a \partial_x^2 v) + b \partial_x v - c v\right](t,x;\beta) & = 0\, \,\,\, \text{in $[0,T)\times D$\,,} \\
v(T,\cdot) &  = g 	\, \,\,\, \text{on $D$} \,.
\end{split}\right.
\end{equation}
Since $v\in C^{1,2}([0,T]\times D)$ and since $v$ satisfies the above PDE, if we apply It\^o's formula then we obtain
\begin{equation} \label{eq bsde}
D(t,T)v(T,X_T^\beta; \beta) = v(t,x;\beta)  + \int_t^T D(t,s) \partial_x v(s,X_s^\beta;\beta) \sigma(s,X_s^\beta; \beta)\,dW_s\,.
\end{equation}
Hence Feynman-Kac representation together with the fact that $v(T,X_T^\beta;\beta)=g(X_T^\beta)$ yields
\begin{equation}\label{eq lin after ito and eq}
 v(t,x;\beta)  = D(t,T)g(X_T^\beta) - \int_t^T D(t,s) \partial_x v(s,X_s^\beta;\beta) \sigma(s,X_s^\beta; \beta)\,dW_s.
\end{equation}
Provided that  
\[
\sup_{s\in[t,T]}\mathbb E [ |D(t,s) \partial_x v(s,X_s^\beta;\beta) \sigma(s,X_s^\beta;\beta)|^2]<\infty\,,
\] 
then the stochastic integral is a martingale. Thus we can consider the Monte-Carlo estimator.
\begin{equation}\label{eq theoretical estimator properties}
\mathcal V^N (t,x;\beta):= \frac{1}{N} \sum_{i=1}^{N} \bigg\{ D^i(t,T)g(X^{\beta,i}_T) - \int_t^T D^i(t,s) \partial_x v(s,X^{\beta,i}_s; \beta) \sigma(s,X^{\beta,i}_s; \beta)\,dW^i_s\bigg\}\,.
\end{equation}

To obtain a control variate we thus need to approximate $\partial_x v$. 
If one used classical approximation techniques to the PDE, 
such as finite difference or finite element methods, 
one would run into the curse of the dimensionality - the very reason one employs Monte-Carlo simulations in the first place. 
Artificial neural networks have been shown to break the curse of dimensionality in specific
situations~\cite{GrohsHornungJentzenvonWurstemberger2018}.
To be more precise, authors in~\cite{Berner_2020,Elbr_chter_2021,jentzen2018proof,grohs2019deep,Hutzenthaler_2020,GrohsHornungJentzenvonWurstemberger2018,kutyniok2020theoretical,gonon2021uniform,reisinger2020rectified} 
have shown that there always exist a deep feed forward neural network and some parameters such that the corresponding neural network can approximate the solution of a linear PDE arbitrarily well in a suitable norm under reasonable assumptions (terminal condition and coefficients can be approximated by neural networks). 
Moreover the number of parameters grows only polynomially in dimension and so there is no curse of dimensionality. 
However, while the papers above construct the network they do not tell us how to find the ``good'' parameters. 
In practice the parameter search still relies on gradient descent-based minimisation over a non-convex landscape.
The application of the deep-network approximation to the solution of the PDE as a martingale control variate is an ideal compromise.


If there is no exact solution to the PDE~\eqref{eq pde}, as would be the
case in any reasonable application, then we will approximate $\partial_x v$ by 
$\mathcal R[\partial_x v]_\theta\in \mathcal{DN}$.  

To obtain an implementable algorithm we discretise the integrals in $\mathcal V_t^{\beta,N,v}$ and take a partition of $[0,T]$ denoted $\pi := \{t=t_0<\cdots<t_{N_\text{steps}}=T\}$, and consider an approximation of~\eqref{eq sde} by $(X_{t_k}^{\beta,\pi})_{t_k\in\pi}$. 
For simplicity we approximate all integrals arising by Riemann sums always taking the left-hand point when approximating the value of the integrand.

The implementable control variate Monte-Carlo estimator is then the form 
\begin{equation}\label{eq cv mc}
\begin{split}
& \mathcal V^{\pi,\theta,\lambda,N}  (t,x;\beta)  :=  \frac{1}{N} \sum_{i=1}^{N} \bigg\{ 
(D^{\pi}(t,T))^ig(X^{\beta,\pi,i}_T) \\
- &\lambda \sum_{k=1}^{N_\text{steps}-1} (D^{\pi}(t,t_k))^i \mathcal R[\partial_x v]_\theta(t_k,X^{\beta,\pi,i}_{t_k}; \beta) \sigma(t_k,X^{\beta,\pi,i}_{t_k}; \beta)\,(W^i_{t_{k+1}} - W^i_{t_{k}}) \bigg\}\,,
\end{split}
\end{equation}
where 
$D^{\pi}(t,T):=e^{-\sum_{k=1}^{N_\text{steps} -1} c(t_k,X^{\beta,\pi}_{t_k} )(t_{k+1} - t_k) }$ 
and $\lambda$ is a free parameter to be chosen (because we discretise and use approximation to the PDE it is expected $\lambda \neq 1$). 
Again, we point out that the only bias of the above estimator comes from the numerical scheme used to solve the forward and backward processes. Nevertheless, $\mathcal R[\partial_x v]_\theta$ does not add any additional bias independently of the choice $\theta$. 
We will discuss possible approximation strategies for approximating $\partial_x v$ with $\mathcal R [\partial_x v]_\theta$ in the following section. 

In this section we have actually derived an explicit form of the Martingale representation (see e.g.~\cite[Th. 14.5.1]{cohen2015stochastic}) of $D(t,T)g(X_T^\beta)$ in terms of the solution of the PDE associated to the process $X^\beta$, which is given as the solution to~\eqref{eq sde}. In Appendix~\ref{sec martingale cv} we provide a more general framework to build a low-variance Monte Carlo estimator $\mathcal V^N_t$ for any (possibly non-Markovian) $\mathcal F^W$-adapted process $X^\beta$.

\subsection{Unbiased Parametric PDE approximation}\label{sec unbiased pde solver}
After having trained the networks $\mathcal R[\partial_x v]_{\theta}$ (using any of Algorithms~\ref{alg PDE cond expec}, ~\ref{alg PDE prob repr iterative training}, ~\ref{alg empirical risk minimisation}, ~\ref{alg empirical corr maximisation} that we will introduce in Section~\ref{sec deep pde solvers}) and $\mathcal R[v]_{\eta}$ (using any of Algorithms~\ref{alg PDE cond expec}, ~\ref{alg PDE prob repr iterative training}) that approximate $v, \partial_x v$ one then has two options to approximate $v(t,x_t;\beta)$
\begin{enumerate}[i)]
 \item Directly with $\mathcal R[v]_{\eta}(t, x_t; \beta)$ if Algorithms~\ref{alg PDE cond expec} or \ref{alg PDE prob repr iterative training} were used, which will introduce some approximation bias.	
 \item By combining $\mathcal R[\partial_x v]_{\theta}$ with the Monte Carlo approximation of $v(t,x_t;\beta)$ using~\eqref{eq cv mc}, which will yield an unbiased estimator of $v(t,x_t;\beta)$. The complete method is stated as Algorithm~\ref{alg unbiased parametric pde solver}.
\end{enumerate}

\begin{algorithm}
\caption{Unbiased parametric PDE solver}
\label{alg unbiased parametric pde solver}
\begin{algorithmic}
\STATE{Input: $t,x,\beta$ where $t\in\pi$.}
\STATE{Initialisation: $\theta$, $N_{\text{trn}}$}
\FOR{$i:1:N_{\text{trn}}$}
\STATE{ 
Generate  samples 
$(x_t^{\beta,\pi,i})_{t\in\pi}$ by using numerical SDE solver on \eqref{eq sde}.
}
\ENDFOR
\STATE{Find the optimal weights $\theta^{*,N_{\text{trn}}}$ of $\mathcal R[\partial_x v]_{\theta}$ using one of Algorithms~\ref{alg PDE cond expec}, ~\ref{alg PDE prob repr iterative training}, ~\ref{alg empirical risk minimisation}, ~\ref{alg empirical corr maximisation}.
 }
\RETURN $\mathcal V^{\beta,\pi,\theta^*,\lambda,N}_{t,T}$ as defined in ~\eqref{eq cv mc} with $\theta$ replaced by $\theta^{*,N_{\text{trn}}}$.
\end{algorithmic}
\end{algorithm}

\section{Deep PDE solvers}\label{sec deep pde solvers}

In this section we propose two algorithms that learn the PDE solution (or its gradient) and then use it to build control variate using~\eqref{eq cv mc}. 
We also include in the Appendix~\ref{sec appendix martingale control variate deep solvers} an additional algorithm to solve such linear PDEs using deep neural networks.  

\subsection{Projection solver}
Before we proceed further we recall a well known property of conditional expectations, for proof see e.g.~\cite[Ch.3 Th. 14]{krylov2002introduction}.
\begin{theorem}\label{th conditional}
Let $\mathcal X\in L^2(\mathcal F)$. Let $\mathcal G \subset \mathcal F$ be a sub $\sigma$-algebra. There exists a random variable $Y\in L^2(\mathcal G)$ such that 
\[
\mathbb E[| \mathcal X - \mathcal Y |^2] = \inf_{\eta \in L^2(\mathcal G)} \mathbb E[| \mathcal X- \eta  |^2].
\] 	
The minimiser, $\mathcal Y$, is unique and is given  by $\mathcal Y=\mathbb E[\mathcal X | \mathcal G]$.
\end{theorem}
The theorem tell us that conditional expectation is an orthogonal projection of a random variable $X$ onto 
$L^2(\mathcal G)$. 
Instead of working directly with~\eqref{eq pde} we work with its probabilistic representation~\eqref{eq bsde}. To formulate the learning task, 
we replace $\mathcal{X}$ by $D(t,T)g((X_T^\beta))$
so that $v(t, X_t^\beta;\beta) = \mathbb{E}[\mathcal{X} | X_t^\beta]$. Hence, by Theorem \eqref{th conditional},
\[
\mathbb E[|  \mathcal X   - v(t,X_{t}^\beta;\beta) |^2] = \inf_{\eta \in L^2(\sigma(X_t^\beta))} \mathbb E[| \mathcal X - \eta  |^2]
\]
and we know that for a fixed $t$ the random variable which minimises the mean
square error is a function of $X_t$.
But by the Doob--Dynkin Lemma~\cite[Th. 1.3.12]{cohen2015stochastic} 
we know that every $\eta \in L^2(\sigma(X_t))$ can be expressed as 
$\eta = h_t(X_t^\beta)$ for some appropriate measurable $h_t$.
For the practical algorithm we restrict the search for the function $h_t$ to the class that can be expressed as deep neural networks $\mathcal {DN}$. 
Hence we consider a family of functions $\mathcal R_\theta \in \mathcal {DN}$ and set learning task as
\begin{equation} \label{eq loss bel}
\theta^{*} := \argmin_{\theta} \mathbb E_\beta\left[ \mathbb E_{(X_t^{\beta,\pi})_{t\in\pi}}\left[ \sum_{k=0}^{N_{\text{steps}}} \left( D(t_k, T)g(X_T^{\beta,\pi}) - \mathcal R[v]_{\theta_{t_k}}(X_{t_k}^{\beta,\pi}; \beta) \right)^2\right]\right].
\end{equation}
The inner expectation in~\eqref{eq loss bel} is taken across all paths generated using numerical scheme on~\eqref{eq sde} for a fixed $\beta$ and it allows to solve the PDE~\eqref{eq pde} for such $\beta$.
The outer expectation is taken on $\beta$ for which the distribution is fixed beforehand (e.g. uniform on $B$ if it is compact), thus allowing the algorithm to find the optimal neural network weights $\theta^*$ to solve the parametric family of PDEs~\eqref{eq pde}. Automatic differentiation is used to approximate $\partial_x v$. Algorithm~\ref{alg PDE cond expec} describes the method. 
\begin{algorithm}
\caption{Projection solver}
\label{alg PDE cond expec}
\begin{algorithmic}
\STATE{Initialisation: $\theta$, $N_{\text{trn}}$, distribution of $\beta$.}
\FOR{$i:1:N_{\text{trn}}$}
\STATE{ generate  samples 
$(x_t^{\beta,\pi,i})_{t\in\pi}$ by using numerical SDE solver on \eqref{eq sde} and sampling from the distribution of $\beta$.
} 
\ENDFOR
\STATE{Use SGD to find $\theta^{*,N_{\text{trn}}}$ where

 \[
 \theta^{*,N_{\text{trn}}}= \argmin_{\theta} \mathbb E^{\mathbb P^{N_{\text{trn}}}} \left[ \sum_{k=0}^{N_{\text{steps}}-1} \left( D(t_k, T)g(X^{\beta,\pi}_T) - \mathcal R[v]_{\theta_{t_k}}(X^{\beta,\pi}_{t_k}; \beta) \right)^2 \right]
 \]
 Where $\mathbb E^{\mathbb P^{N_{\text{trn}}}}$ denotes the empirical mean. 
  }
\STATE{Automatic differentiation applied to $\mathcal R[v]_{\theta_{t_k}^*}(x^{\beta,\pi}_{t_k}; \beta)$ can be used to approximate $\partial_x v$.
}
\RETURN $\theta^{*,N_{\text{trn}}}$.
\end{algorithmic}
\end{algorithm}

\subsection{Probabilistic representation based on Backward SDE}
\label{sec prob repr}

Instead of working directly with~\eqref{eq pde} we work with its probabilistic representation~\eqref{eq bsde} and view it as a
BSDE.
To formulate the learning task based on this we 
recall the time-grid $\pi$ 
so that we can write it recursively as
\[
\begin{split} 
\label{eq representation delta}
& v(t_{N_\text{steps}},X_{t_{N_\text{steps}}}^{\beta}; \beta) =  g(X_{t_{N_\text{steps}}}^{\beta})\,,\\
D(t,t_{m+1}) & v(t_{m+1},X_{t_{m+1}}^{\beta}; \beta) = D(t,t_{m})v(t_m,X_{t_m}^{\beta}; \beta)\,\,\,\,\, \\
& + \int_{t_m}^{t_{m+1}} D(t,s)\partial_x v(s,X_s^{\beta}; \beta) \sigma(s,X_s^\beta; \beta)\,dW_s\,\,\,\text{for}\,\,\,m=0,1,\ldots,N_{\text{steps}}-1\,.\\
\end{split}
\]

Next
consider deep network approximations for each time step in $\pi$ and for both the solution of ~\eqref{eq pde} and its gradient.  
\[
\mathcal R[v]_{\eta_m}(x;\beta) \approx v (t_m,x; \beta)\,,\,\,\,\, t_m\in\pi\,,\,\,\, x \in \mathbb R^d\,
\]
and
\[
\mathcal R[\partial_x v]_{\theta_m}(x; \beta) \approx \partial_x v(t_m,x; \beta)\,,\,\,\,\, t_m\in\pi\,,\,\,\, x \in \mathbb R^d\,.
\]
Approximation depends on weights $\eta_m \in \mathbb R^{k_\eta}$, $\theta_m \in \mathbb R^{k_\theta}$. 
We then set the learning task as
\begin{equation} \label{eq loss function BSDE solver}
\begin{split}
 (\eta^*, \theta^*) := & \argmin_{(\eta,\theta)} \mathbb{E}_{\beta, X^{\beta}} 
 \bigg[ \left|g(X_{t_{N_\text{steps}}}^{\beta,\pi}) -  \mathcal R[v]_{\eta_{N_\text{steps}}}(X_{t_{N_\text{steps}}}^{\beta,\pi}) \right|^2\\
& \qquad \qquad + \frac{1}{N_\text{steps}}\sum_{m=0}^{N_\text{steps}-1}|\mathcal E^{(\eta,\theta)}_{m+1}|^2
\bigg]\,,\\
\mathcal E^{(\eta,\theta)}_{m+1} 
:= & D(t,t_{m+1}) \mathcal R[v]_{\eta_{m+1}}(X_{t_{m+1}}^{\beta,\pi}; \beta) -  D(t,t_{m}) \mathcal R[v]_{\eta_m}(X^{\beta,\pi}; \beta) \\
& \qquad \qquad  - D(t,t_{m}) \mathcal R[\partial_x v]_{\theta_m}(X_{t_m}^{\beta,\pi}; \beta)\sigma(t_m,X_{t_m}^{\beta, \pi}; \beta) \Delta W_{t_{m+1}}\,,
\end{split}
\end{equation}
where
\[
\eta = \{\eta_0, \ldots, \eta_{t_{N_\text{steps}}}\}, \, \, \theta = \{\theta_0, \ldots, \theta_{t_{N_\text{steps}}}\}\, .
\]
The complete learning method is stated as Algorithm~\ref{alg PDE prob repr iterative training}, where
we split the optimisation \eqref{eq loss function BSDE solver} in several optimisation problems, one per time step: 
learning the weights $\theta_m$ or $\eta_m$ at a certain time step $t_m<t_{N_\text{steps}}$ only requires 
knowing the weights $\eta_{m+1}$. 
At $m=N_{\text{steps}}$, learning the weights $\eta_{N_{\text{steps}}}$ only
requires the terminal condition $g$.  Note 
that the algorithm assumes that adjacent networks in time will be similar, and therefore we 
initialise $\eta_m$ and $\theta_m$ by $\eta_{m+1}^*$ and $\theta_{m+1}^*$.

\begin{algorithm}
\caption{Martingale representation solver, iterative}
\label{alg PDE prob repr iterative training}
\begin{algorithmic}
\STATE{Initialisation: $N_{\text{trn}}$}
\FOR{$i:1:N_{\text{trn}}$}
\STATE{ generate  samples 
$(x_t^{\beta,\pi,i})_{t\in\pi}$ by using numerical SDE solver on \eqref{eq sde} and sampling from the distribution of $\beta$.} 
\ENDFOR
\STATE{Initialisation: $\eta_{N_{\text{steps}}}$}
\STATE{Find $\eta_{N_{\text{steps}}}^{*,N_{\text{trn}}}$ using SGD where
\[
\begin{split}
\eta_{N_{\text{steps}}}^{*,N_{\text{trn}}} := & \argmin_{\eta} \frac{1}{N_{\text{trn}}} \sum_{i=1}^{N_{\text{trn}}} \left|g(x^{\beta,\pi,i}_{t_{N_\text{steps}}}) - \mathcal R[v]_{\eta_{N_\text{steps}}}(x^{\beta,\pi,i}_{t_{N_\text{steps}}};\beta) \right|^2 \,
\end{split}
\]
}
\FOR{$m:N_{\text{steps}}-1:0:-1$}
\STATE{Initialise $(\theta_m, \eta_m) = (\theta_{m+1}^{*,N_{\text{trn}}}, \eta_{m+1}^{*,N_{\text{trn}}})$}
\STATE{Find $(\theta_m^{*,N_{\text{trn}}}, \eta_m^{*,N_{\text{trn}}})$ using SGD where
\[
(\theta_m^{*,N_{\text{trn}}}, \eta_m^{*,N_{\text{trn}}}) := \argmin_{(\eta_m,\theta_m)} \frac{1}{N_{\text{trn}}} \sum_{i=1}^{N_{\text{trn}}} \left| \mathcal E^{\beta, \pi,i,(\eta,\theta)}_{m+1}\right|^2 
\]
where 
\[
\begin{split}
 \mathcal E^{\pi,i,(\eta,\theta)}_{m+1} 
:= & D^{\pi,i}(t,t_{m+1})\mathcal R[v]_{\eta_{m+1}}(x^{\beta,\pi,i}_{t_{m+1}};\beta) -  D^{\pi,i}(t,t_{m}) \mathcal R[v]_{\eta_m}(x^{\beta,\pi,i}_{t_m};\beta) \\
& \qquad \qquad  - D^{\pi,i}(t,t_{m}) \mathcal R[\partial_x v]_{\theta_m}(x^{\beta,\pi,i}_{t_m};\beta)\sigma(t_m,x^{\beta,\pi,i}_{t_m}; \beta) \Delta W^i_{t_{m+1}}\,.
\end{split}
\]
}
\ENDFOR
\RETURN $(\theta_m^{*,N_{\text{trn}}}, \eta_m^{*,N_{\text{trn}}})$ for all $m=0,1,\ldots,N_{\text{steps}}$.
\end{algorithmic}
\end{algorithm}

%

\subsection{Martingale Control Variate deep solvers}\label{sec martingale cv deep solvers}
So far, the presented methodology to obtain the control variate consists on first learning the solution of the PDE and more importantly its gradient (Algorithms~\ref{alg PDE cond expec},~\ref{alg PDE prob repr iterative training}) which is then plugged in~\eqref{eq cv mc}. Alternatively, one can directly use the variance of~\eqref{eq cv mc} as the loss function to be optimised in order to learn the control variate. We expand this idea and design two additional algorithms.

Recall definition of $\mathcal V^{\beta,\pi,\theta,\lambda,N}_{t,T}$ 
given by~\eqref{eq cv mc}. 
From~\eqref{eq theoretical estimator properties} we know that the 
theoretical control variate Monte-Carlo estimator has zero variance 
and so it is natural to set-up a learning task which aims to learn
the network weights $\theta$ in a way which minimises said variance: 
\[
 \theta^{\star,\text{var}} :=  \argmin_\theta \mathbb{V}\text{ar} \Big[\mathcal V^{\beta,\pi,\theta,\lambda,N}_{t,T} \Big]\,.
\]
Setting $\lambda=1$, the learning task is stated as Algorithm~\ref{alg empirical risk minimisation}.

\begin{algorithm}
\caption{Martingale control variates solver: Empirical variance minimisation}
\label{alg empirical risk minimisation}
\begin{algorithmic}
\STATE{Initialisation: $\theta$, $N_{\text{trn}}$}
\FOR{$i:1:N_{\text{trn}}$}
\STATE{ 
generate  samples 
$(x_t^{\beta,\pi,i})_{t\in\pi}$ by using numerical SDE solver on \eqref{eq sde} and sampling from the distribution of $\beta$.
}
\ENDFOR
\STATE{Find $\theta^{*,N_{\text{trn}}}$ where 
\[
\theta^{*, N_{\text{trn}}} := \argmin_{\theta} \mathbb V\text{ar}^{N_{\text{trn}}}  \left[\mathcal V^{\beta,\pi,\theta,\lambda,N_{\text{trn}}}_{t,T} \right],
\]
where $\mathbb Var^{N_{\text{trn}}}$ denotes the empirical variance, and $\mathcal V^{\beta,\pi,\theta,\lambda,N}_{t,T}$ is obtained from~\eqref{eq cv mc}.
 }
\RETURN $\theta^{*,N_{\text{trn}}}$.
\end{algorithmic}
\end{algorithm}

We include a second similar Algorithm in Appendix~\ref{sec appendix martingale control variate deep solvers}.

\section{Examples and experiments}
\label{sec examples}

\subsection{Options in Black--Scholes model on $d>1$ assets}
\label{sec examples model}
Take a $d$-dimensional Wiener process $W$. 
We assume that we are given a symmetric, positive-definite matrix 
(covariance matrix) $\Sigma$ and a lower triangular matrix $C$ s.t.
$\Sigma=CC^*$.
For such a positive-definite $\Sigma$ we can always use Cholesky decomposition to find $C$. 
The risky assets will have volatilities given by $\sigma^i$. 
We will (abusing notation) write $\sigma^{ij} := \sigma^i C^{ij}$, when we don't
need to separate the volatility of a single asset from correlations.
The risky assets under the risk-neutral measure are then given by 
\begin{equation} \label{eq BS SDE}
dS^i_t = rS^i_t \, dt + \sigma^i S^i_t \sum_j C^{ij} dW^j_t\,.
\end{equation}
All sums will be from $1$ to $d$ unless indicated otherwise.
Note that the SDE can be simulated exactly since
\[
S^i_{t_{n+1}} = S^i_{t_n}\exp\left(\left(r-\frac12  \sum_j (\sigma^{ij})^2\right)(t_{n+1}-t_n)+\sum_j \sigma^{ij}(W^j_{t_{n+1}}-W^j_{t_n}) \right)\,.
\]
The associated PDE is (with $a^{ij} :=  \sum_k \sigma^{ik}\sigma^{jk}$) 
\[
\partial_t v(t,S) + \frac{1}{2}\sum_{i,j} a^{ij} S^i S^j \partial_{x_i x_j}v(t,S)  + r \sum_i S^i \partial_{S^i} v(t,S) - r v(t,S) = 0\,,
\]
for $(t,S) \in [0,T)\times (\mathbb R^+)^d$ together with the
terminal condition $v(T,S) = g(S)$ for $S\in \mathbb (\mathbb R^+)^d$.

%
%
%

\subsection{Deep Learning setting}
\label{sec experiment deep learning setting}
In this subsection we describe the neural networks  used in the four proposed algorithms
as well as the training setting, in the specific situation where we have an options problem in 
Black-Scholes model on $d>1$ assets. 

%

Learning algorithms~\ref{alg PDE prob repr iterative training}, \ref{alg empirical risk minimisation}
and~\ref{alg empirical corr maximisation}
share the same underlying fully connected artificial network
which will be different for different $t_k$, $k=0,1,\ldots,N_{\text{steps}}-1$.
At each time-step we use a fully connected artificial neural network 
denoted $\mathcal R[\cdot]_{\theta_k} \in \mathcal{DN}$. 
The choice of the number of layers and network width is motivated by empirical results on 
different possible architectures applied on a short-lived options problem. 
We present the results of this study in Appendix~\ref{sec experiments empirical net diag}. 
The architecture is similar to that proposed in~\cite{beck2018solving}.

At each time step the network consists of four layers: 
one $d$-dimensional input layer, 
two $(d+20)$-dimensional hidden layers, and one 
output layer. 
The output layer is one dimensional if the network is approximation for $v$
and $d$-dimensional if the network is an approximation for $\partial_x v$.
The non-linear activation function used on the hidden layers is the the linear rectifier \texttt{relu}. 
In all experiments except for Algorithm~\ref{alg PDE prob repr iterative training}
for the basket options problem 
we used batch normalisation~\cite{Ioffe2015normalization} on the input of each network, 
just before the two nonlinear activation functions in front of the hidden layers, 
and also after the last linear transformation.

The networks' optimal parameters are approximated by the Adam optimiser \cite{diedrik2017adam} on the loss function 
specific for each method. 
Each parameter update (i.e. one step of the optimiser) is calculated on a 
batch of $5\cdot 10^3$ paths $(x_{t_n}^i)_{n=0}^{N_\text{steps}}$ obtained
by simulating the SDE. 
We take the necessary number of training steps until the stopping criteria defined below is met, with a learning rate of $10^{-3}$ during the first $10^4$ iterations, decreased to $10^{-4}$ afterwards. 


During training of any of the algorithms, the loss value at each iteration is kept. 
A model is assumed to be trained if the 
difference between the loss averages of the two last consecutive windows of length 100 is less than a certain $\epsilon$.


\subsection{Evaluating variance reduction}
We use the specified network architectures to assess the variance reduction in several 
examples below. After training the models in each particular example, they are evaluated as follows:
\begin{enumerate}[i)]
\item We calculate $N_{\text{MC}} = 10$ times the Monte Carlo estimate $\overline{\Xi_T}:= \frac1{N_{\text{in}}}\sum_{i=1}^{N_{\text{in}}}\Xi_T^i$ and the Monte Carlo
with control variate estimate
$\bar{\mathcal V}^{\pi,\theta, \lambda, N_{\text{steps}}}_{t,T} = \frac1{N_{\text{in}}}\sum_{i=1}^{N_{\text{in}}} \mathcal V^{\pi,\theta,\lambda,N_{\text{steps}}, i}_{t,T}$ using 
$N_{\text{in}}=10^6$ Monte Carlo samples.
\item From Central Limit Theorem, as $N_{\text{in}}$ increases the standardised estimators converge in distribution to 
the Normal. Therefore, a 95\% confidence interval of the variance of the estimator is given by 
\[
\left[ \frac{(N_{\text{MC}}-1)S^2}{\chi_{1-\alpha/2, N_{\text{MC}}-1}}, \frac{(N_{\text{MC}}-1)S^2}{\chi_{\alpha/2,N_{\text{MC}}-1}}\right]
\]
where $S$ is the sample variance of the $N_{\text{MC}}$ controlled estimators $\bar{\mathcal V}^{\pi,\theta,\lambda,N_{\text{steps}}}_{t,T}$, 
and $\alpha = 0.05$. These are calculated for both the Monte Carlo estimate and the Monte Carlo with control variate estimate. 
\item We use the $N_{\text{MC}} \cdot N_{\text{in}} = 10^7$ generated samples $\Xi_T^i$ and $\mathcal V^{\pi,\theta,\lambda,N_{\text{steps}}, i}_{t,T}$ 
to calculate and compare the empirical variances $\tilde{\sigma}^2_{\Xi_T}$ and $\tilde{\sigma}^2_{\mathcal V^{\pi,\theta,\lambda,N_{\text{steps}}, i}_{t,T}}$.
\item The number of optimizer steps and equivalently number of random paths generated for training provide a cost measure of the proposed algorithms. 
\item We evaluate the variance reduction if we use the trained models to create control variates for options in Black-Scholes models 
with different volatilities than the one used to train our models. 
\end{enumerate}



\begin{example}[Low dimensional problem with explicit solution]
\label{ex marg}
We consider exchange option on two assets. 
In this case the exact price is given by the Margrabe formula. 
We take $d=2$, $S^i_0 = 100$, $r=5\%$, $\sigma^i = 30\%$, $\Sigma^{ii} = 1$, $\Sigma^{ij} = 0$ for $i\neq j$.	
The payoff is
\[
g(S) = g(S^{(1)},S^{(2)}) := \max\left(0,S^{(1)} - S^{(2)}\right)\,.
\]
From Margrabe's formula we know that
\[
v(0,S) = \text{BlackScholes}\left(\text{risky price}=\frac{S^{(1)}}{S^{(2)}}, \text{strike}=1, T, r, \bar \sigma \right)\,,
\]
where $\bar \sigma := \sqrt{\left(\sigma^{11} - \sigma^{21}  \right)^2
+\left(\sigma^{22}-\sigma^{12}\right)^2}$\,.

We organise the experiment as follows: 
We train our models with batches of 5,000 random paths $(s_{t_n}^i)_{n=0}^{N_\text{steps}}$ sampled from the SDE~\ref{eq BS SDE},
where $N_\text{steps}=50$. The assets' initial values $s_{t_0}^i$ are sampled from a lognormal distribution 
\[
X \sim \exp ((\mu-0.5\sigma^2)\tau + \sigma\sqrt{\tau}\xi),
\]
where $\xi\sim \mathcal{N}(0,1), \mu=0.08, \tau=0.1$. The existence of an explicit solution allows to 
build a control variate of the form 
\eqref{eq cv mc} using the known exact solution to obtain $\partial_x v$.
For a fixed number of time steps $N_\text{steps}$ this provides 
an upper bound on the variance reduction an artificial neural network approximation of $\partial_x v$ can achieve.

We follow the evaluation framework to evaluate the model, simulating $N_{\text{MC}} \cdot N_{\text{in}}$ paths by 
simulating~\eqref{eq BS SDE} with constant $(S^1_0, S^2_0)^i = (1,1)$. 
We report the following results:
\begin{enumerate}[i)]
\item Table~\ref{table results empirical variance exchange 2d} provides the empirical variances calculated over $10^6$ generated Monte Carlo samples and their corresponding control variates. 
The variance reduction measure indicates the quality of each control variate method. The 
variance reduction using the control 
variate given by Margrabe's formula provides a benchmark for our methods. Table~\ref{table results empirical variance exchange 2d}
also provides the cost of training for each method, given by the number of optimiser iterations performed before 
hitting the stopping criteria, defined defined before with $\epsilon = 5\times 10^{-6}$. 
We add an additional row with the control variate built using automatic differentiation on the network parametrised using the Deep Galerkin Method~\cite{sirignano2017dgm}. The DGM attempts to find the optimal parameters of the network satisfying the PDE on a pre-determined time and space domain. 
In contrast to our algorithms, the DGM method is not restricted to learn the solution of the PDE on the paths built from the probabilistic representation of the PDE. However, this is what is precisely enhancing the performance of our methods in terms of variance reduction, since they are specifically learning an approximation to the solution of the PDE and its gradient such that the resulting control variate will yield a low-variance Monte Carlo estimator. 
\item Table~\ref{table results conf interval exchange 2d} provides the confidence intervals for the variances and  
of the Monte Carlo estimator, and the Monte Carlo estimator with control variate assuming these 
are calculated on $10^6$ random paths. 
Moreover, we add the confidence interval of the variance of the Monte Carlo estimator calculated over $N_{in}$ antithetic paths where the first   $N_{in}/2$ Brownian paths generated using $(Z_i)_{i=1,...,N_{steps}}$ samples from a normal and the second half of the Brownian paths are generated using the antithetic samples $(-Z_i)_{i=1,...,N_{steps}}$. 
See~\cite[Section 4.2]{belomestny2017variance} for more details. All the proposed algorithms in this paper outperform the Monte Carlo estimator and the Monte Carlo estimator with antithetic paths; compared to the latter, our algorithms produce unbiased estimators with variances that two orders of magnitude less.
\item Figure~\ref{fig var red it eps exchange 2dim} studies the iterative training 
for the BSDE solver. As it has been observed before, this type of training does not allow us to study the overall loss function 
as the number of training steps increases. 
Therefore we train the same model four times for different values of $\epsilon$ 
between $0.01$ and $5\times 10^{-6}$ and we study the number of iterations necessary to meet the stopping criteria
defined by $\epsilon$, the variance reduction once the stopping criteria is met, and the relationship between the 
number of iterations and the variance reduction. Note that the variance reduction stabilises for $\epsilon<10^{-5}$. 
Moreover, the number of iterations necessary to meet the stopping criteria increases exponentially as $\epsilon$ decreases,
and therefore for our results printed in Tables~\ref{table results empirical variance exchange 2d} and~\ref{table results conf interval exchange 2d} we employ $\epsilon = 5\times 10^{-6}$.
\item Figure~\ref{fig var reduction sigmas exchange 2dim} displays the variance reduction after using the trained
models on several Black Scholes problem with exchange options but with values of $\sigma$ other than $0.3$ which was
the one used for training. We see that the various algorithms work similarly well in this case (not taking
training cost into account). 
We note that the variance reduction is close to the theoretical maximum which
is restricted by time discretisation. 
Finally we see that the variance reduction is still significant even when the
neural network was trained with different model parameter (in our case
volatility in the option pricing example). The labels of Figure~\ref{fig var reduction sigmas exchange 2dim} 
can be read as follows:
\begin{enumerate}[i)]
\item \textit{MC + CV Corr op}: Monte-Carlo estimate with Deep Learning-based Control Variate built using Algorithm~\ref{alg empirical corr maximisation}.
\item  \textit{MC + CV Var op}: Monte-Carlo estimate with Deep Learning-based Control Variate built using Algorithm~\ref{alg empirical risk minimisation}.
\item  \textit{MC + CV BSDE solver}: Monte-Carlo estimate with Deep Learning-based Control Variate built using Algorithm~\ref{alg PDE prob repr iterative training}.
\item \textit{MC + CV Margrabe}: Monte-Carlo estimate with Control Variate using analytical solution for this problem given by Margrabe formula.
\end{enumerate}

\end{enumerate}

\begin{table}
\begin{tabular}{|l||c|c|c|c|} 
\hline 
Method & Emp. Var. & Var. Red. Fact. & Train. Paths & Opt. Steps \\
\hline 
\hline 
Monte Carlo & $3.16 \times 10^{-2}$ & - & - & -\\
\hline 
Algorithm~\ref{alg PDE cond expec} & $2.47\times 10^{-4}$  & $127.7$ & $38 \times 10^6$ & $7\, 600$  \\ 
\hline
Algorithm~\ref{alg PDE prob repr iterative training} & $2.59\times 10^{-4}$ & $121.98$ & $6.945\times 10^6$ & $1380$  \\ 
\hline
Algorithm~\ref{alg empirical risk minimisation} & $2.39 \times 10^{-4}$ & $132.28$ &  $36.055\times 10^6$ & $7211$\\ 
\hline 
Algorithm~\ref{alg empirical corr maximisation} &  $2.40\times 10^{-4}$ & $131.53$ & $45.61\times 10^6$ & $9122$ \\ 
\hline
MC + CV Margrabe & $2.12\times 10^{-4}$  & $149.19$ & - & - \\  
\hline \hline
MC + CV DGM \cite{sirignano2017dgm} & $1.22\times 10^{-3}$  & $25.8$ & - & - \\  
\hline
\end{tabular} 
\caption{Results on exchange option problem on two assets, Example~\ref{ex marg}. Empirical Variance and variance reduction factor}
\label{table results empirical variance exchange 2d} 	
\end{table}

\begin{table}
\begin{adjustbox}{max width=\textwidth}
\begin{tabular}{|l||c|c|} 
\hline 
Method & Confidence Interval Variance & Confidence Interval Estimator  \\ 
\hline 
\hline
Monte Carlo & $[2.36\times 10^{-6}, 4.15\times 10^{-6}]$ & $[0.1187,0.1195]$ \\
\hline 
Monte Carlo + antithetic paths & $[1.15\times 10^{-6}, 2.02\times 10^{-6}]$ & $[0.1191,0.1195]$ \\
\hline 
Algorithm~\ref{alg PDE cond expec} & $[4.13\times 10^{-9}, 1.09\times 10^{-8}]$ & $[0.11919,0.11926]$ \\
\hline
Algorithm~\ref{alg PDE prob repr iterative training} & $[4.12\times 10^{-9}, 1.09\times 10^{-8}]$ & $[0.11919,0.11925]$ \\
\hline
Algorithm~\ref{alg empirical risk minimisation} & $[4.32\times 10^{-9}, 1.14\times 10^{-8}]$ & $[0.11919,0.11926]$ \\
\hline 
Algorithm~\ref{alg empirical corr maximisation} & $[2.30\times 10^{-9}, 6.12\times 10^{-8}]$ & $[0.11920,0.11924]$ \\
\hline
MC + CV Margrabe & $[3.10\times 10^{-9}, 8.23\times 10^{-9}]$  & $[0.11919, 0.11925]$  \\  
\hline \hline
MC + CV DGM \cite{sirignano2017dgm} & $[3.10\times 10^{-9}, 8.23\times 10^{-9}]$  & $[0.11919, 0.11925]$  \\ 
\hline

\end{tabular} 
\end{adjustbox}
\caption{Results on exchange option problem on two assets, Example~\ref{ex marg}.}
\label{table results conf interval exchange 2d} 
\end{table}


\begin{figure}[H]
\includegraphics[width=0.5\linewidth]{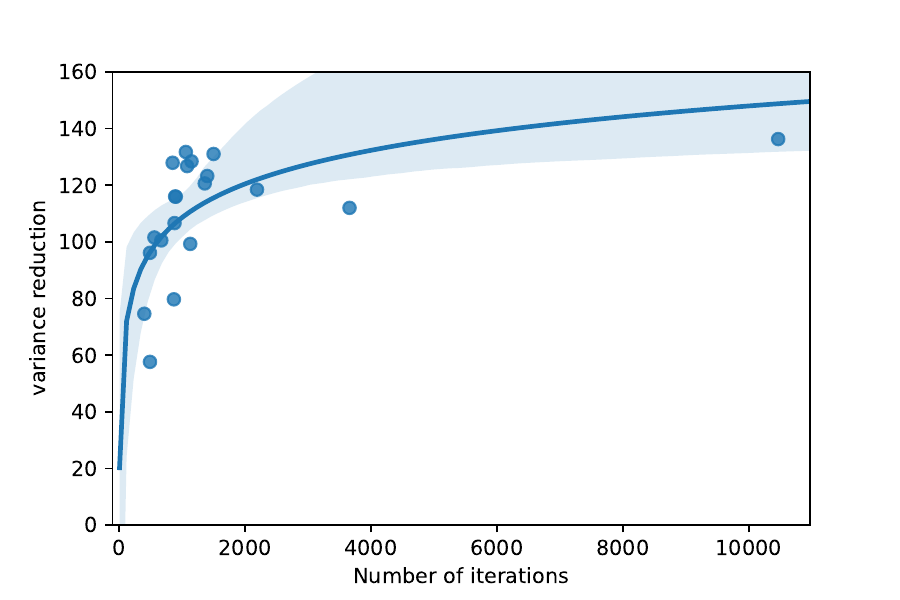}\includegraphics[width=0.5\linewidth]{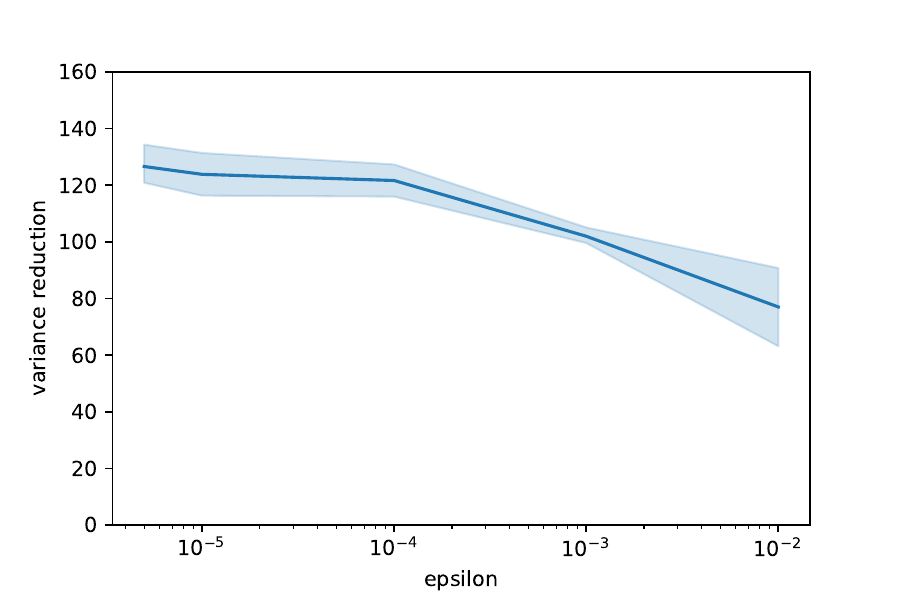}
\caption{Left: Variance reduction in terms of number of optimiser iterations. Right: Variance reduction in terms of epsilon. Both are for Example~\ref{ex marg}
and Algorithm~\ref{alg PDE prob repr iterative training}.}
\label{fig var red it eps exchange 2dim}
\end{figure}

\begin{figure}[H]
\includegraphics[width=0.5\linewidth]{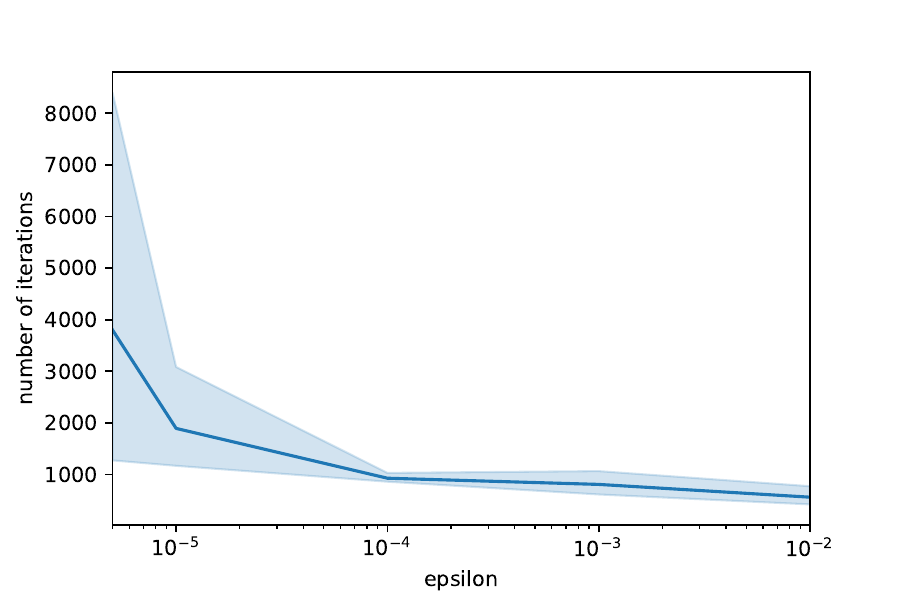}
\caption{Number of optimiser iterations in terms of epsilon 
for Example~\ref{ex marg}
and Algorithm~\ref{alg PDE prob repr iterative training}.}
\label{fig it eps exchange 2dim}
\end{figure}

\begin{figure}
\includegraphics[width=0.8\linewidth]{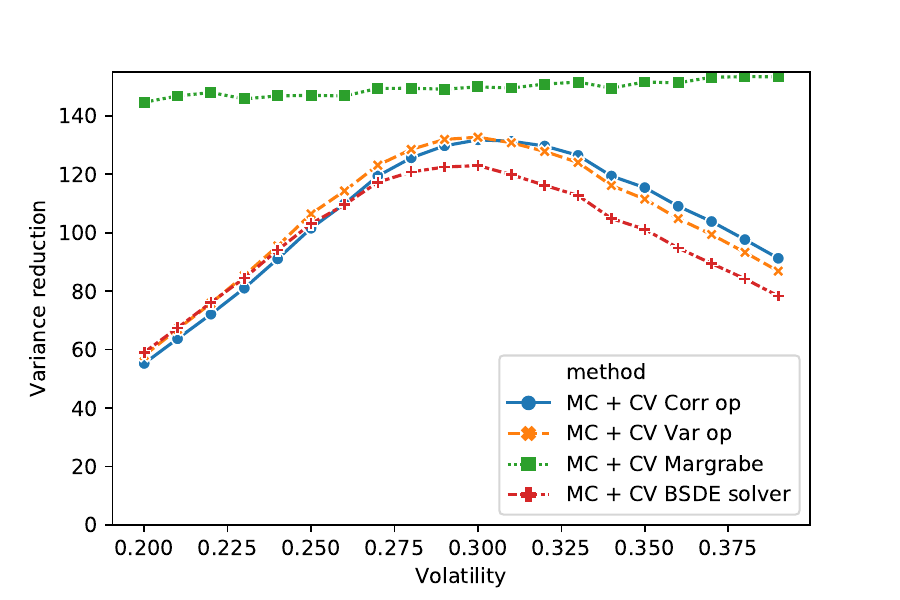}
\caption{Variance reduction achieved by network trained with $\sigma=0.3$ but then applied 
in situations where $\sigma \in [0.2,0.4]$.
We can see that the significant variance reduction is achieved by a neural network
that was trained with ``incorrect'' $\sigma$.
Note that the ``MC + CV Margbrabe'' displays the optimal variance reduction that
can be achieved by using exact solution to the problem. 
The variance reduction is not infinite even in this case since stochastic integrals are approximated by Riemann sums.}
\label{fig var reduction sigmas exchange 2dim}
\end{figure}
%

\end{example}

\begin{example}[Low-dimensional problem with explicit solution - Approximation of Price using PDE solver compared 
to Control Variate]
\label{ex 2d exchange price approx vs CV}
We consider exchange options on two assets as in Example \ref{ex marg}. We consider algorithm~\ref{alg PDE prob repr iterative training} that can be applied in two different ways:
\begin{enumerate}[i)]
\item It directly approximates the solution of the PDE~\eqref{eq pde} and its gradient in every point.
\item We can use $\partial_x v$ to build the control variate using probabilistic representation of the PDE ~\eqref{eq bsde} 
\end{enumerate}
We compare both applications by calculating the expected error of the $L^2$-error of each of them with respect to the analytical 
solution given by Margrabe formula. From Margrabe's formula we know that
\[
v(0,S) = \text{BlackScholes}\left(\text{risky price}=\frac{S^{(1)}}{S^{(2)}}, \text{strike}=1, T, r, \bar \sigma \right)\,,
\]
Let $\mathcal R[v]_{\eta_0} (x) \approx v(0,x)$ be the Deep Learning approximation of price at any point at initial time, 
calculated using Algorithm~\ref{alg PDE prob repr iterative training}, and $\mathcal R[\partial_x v]_{\theta_m}(x) \approx \partial_x v(t_m,x)$ be the Deep Learning 
approximation of its gradient for every time step in the time discretisation.
The aim of this experiment is to show how even if Algorithm ~\ref{alg PDE prob repr iterative training} numerically converges to a biased approximation of $v(0,x)$ (see Figure~\ref{fig CV vs price} left), it is 
still possible to use  $\mathcal R[\partial_x v]_{\theta_m} (x)$ to build an unbiased Monte-Carlo approximation of $v(0,x)$ with low variance. 

We organise the experiment as follows. 

\begin{enumerate}[i)]
\item  We calculate the expected value of the $L^2$-error of $\mathcal R \eta_0 (x)$ where each component of $x \in \mathbb R^2$ is sampled from a lognormal distribution: 
\[
\mathbb E[|v(0,x) -  \mathcal R[v]_{\eta_0} (x)|^2] \approx \frac{1}{N} \sum_{i=1}^N |v(0,x^i) -  \mathcal R[v]_{\eta_0} (x^i)|^2
\]
\item We calculate the expected value of the $L^2$-error of the Monte-Carlo estimator with control variate where each component of $x \in \mathbb R^2$ is sampled from a lognormal distribution: 
\[
\mathbb E[|v(0,x) - \mathcal V^{\pi,\theta,\lambda,N_{MC}, x}_{0,T}|^2] \approx \frac{1}{N} \sum_{i=1}^N |v(0,x^i) -  \mathcal V^{\pi,\theta,\lambda,N_{MC}, x^i}_{0,T} |^2 \, ,
\]
where $\mathcal V^{\pi,\theta,\lambda,N_{MC},x}_{0,T}$ is given by~\ref{eq cv mc},  and is calculated for different values of Monte Carlo samples.
\item We calculate the expected value of the $L^2$-error of the Monte-Carlo estimator without control variate where each component of $x \in \mathbb R^2$ is sampled from a lognormal distribution:
\[
\mathbb E[|v(0,x) -  \Xi^{\pi,\theta,\lambda,N_{MC},x}_{0,T} |^2] \approx \frac{1}{N} \sum_{i=1}^N |v(0,x^i) -  \Xi^{\pi,\theta,\lambda,N_{MC},x^i}_{0,T} |^2 \, ,
\]
where 
\[
\Xi^{\pi,\theta,\lambda,N_{MC},x}_{0,T} := \frac{1}{N_{MC}} \sum_{j=1}^{N_{MC}} D(t,T)g(X_T^i)
\]
\end{enumerate}

Figure \ref{fig CV vs price} provides one realisation of the described experiment for different Monte-Carlo iterations between 10 and 200. It shows how
in this realisation, 60 Monte-Carlo iterations are enough to build a Monte-Carlo estimator with control variate having lower bias than the solution 
provided by Algorithm~\ref{alg PDE prob repr iterative training}.
\begin{figure}[H]
\includegraphics[width=0.49\linewidth]{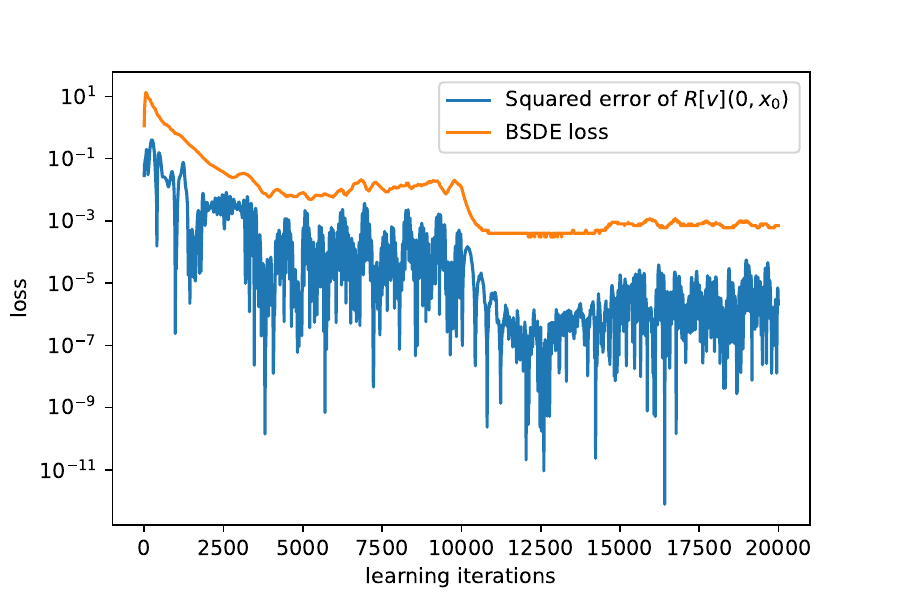}
\includegraphics[width=0.49\linewidth]{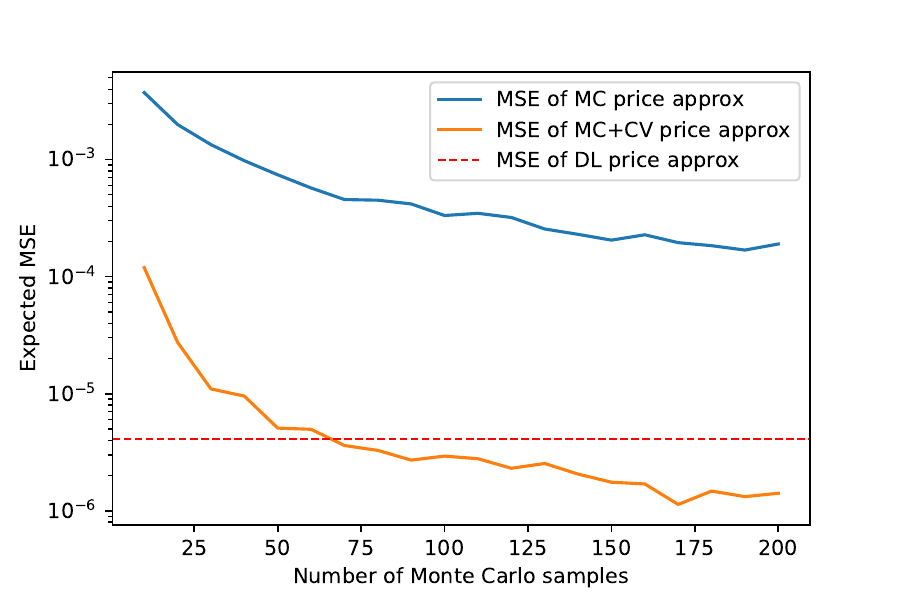}
\caption{Left: Loss of Algorithm~\ref{alg PDE prob repr iterative training} and squared error of $\mathcal R[v](t,x_0)$ in terms of training iterations. Right: Expected MSE of the two different approaches with respect to analytical solution in terms of number 
of Monte Carlo samples}
\label{fig CV vs price}
\end{figure}

\end{example}

\begin{example}[Low-dimensional problem with explicit solution. Training on random values for volatility]
We consider exchange option on two assets. 
In this case the exact price is given by the Margrabe formula. 
The difference with respect to the last example is that now we aim to generalise our model, so that
it can build control variates for different Black-Scholes models.
For this we take $d=2$, $S^i_0 = 100$, $r=0.05$, $\sigma^i \sim \text{Unif}(0.2,0.4)$, $\Sigma^{ii}=1$, $\Sigma^{ij}=0$ for $i\neq j$.

The payoff is
\[
g(S) = g(S^{(1)},S^{(2)}) := \max\left(0,S^{(1)} - S^{(2)}\right)\,.
\]
We organise the experiment as follows: 
for comparison purposes with the BSDE solver from the previous example, we train our model for exactly the same number of iterations, i.e. $1,380$ batches of 5,000 
random paths $(s_{t_n}^i)_{n=0}^{N_\text{steps}}$ sampled from the SDE~\ref{eq BS SDE},
where $N_\text{steps}=50$. The assets' initial values $s_{t_0}^i$ are sampled from a lognormal distribution 
\[
X \sim \exp ((\mu-0.5\sigma^2)\tau + \sigma\sqrt{\tau}\xi),
\]
where $\xi\sim \mathcal{N}(0,1), \mu=0.08, \tau=0.1$. Since now $\sigma$ can take different values, it is
included as input to the networks at each time step. 

The existence of an explicit solution allows to 
build a control variate of the form 
\eqref{eq cv mc} using the known exact solution to obtain $\partial_x v$.
For a fixed number of time steps $N_\text{steps}$ this provides 
an upper bound on the variance reduction an artificial neural network approximation of $\partial_x v$ can achieve. 

Figure~\ref{fig var reduction random sigma} adds the performance of this model to Figure~\ref{fig var reduction sigmas exchange 2dim},
where the variance reduction of the Control Variate is displayed for different values of the volatility between $0.2$ and $0.4$. 

\begin{figure}[H]
\includegraphics[width=0.8\linewidth]{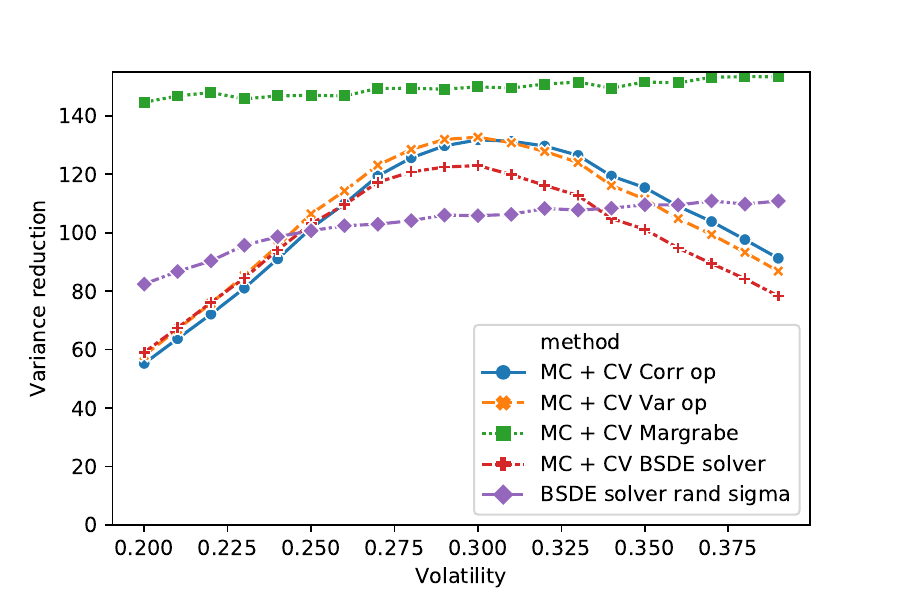}
\caption{Extension of Figure~\ref{fig var reduction sigmas exchange 2dim} with variance reduction 
achieved by training the model on different Black-Scholes models}
\label{fig var reduction random sigma}
\end{figure}
\end{example}

\begin{example}[High-dimensional problem, exchange against average]
\label{ex 100d average exchange}
We extend the previous example to 100 dimensions. 
This example is similar to $\text{EX}_{10E}$ from~\cite{broadie2015risk}. 
We will take $S^i_0 = 100$, $r=5\%$, $\sigma^i = 30\%$, $\Sigma^{ii} = 1$, $\Sigma^{ij} = 0$ for $i\neq j$.	

We will take this to be
\[
g(S):= \max\left(0,S^1 - \frac{1}{d-1}\sum_{i=2}^d S^i\right)\,.
\]

The experiment is organised as follows: we train our models with batches of $5\cdot 10^3$ random paths $(s_{t_n}^i)_{n=0}^{N_\text{steps}}$ sampled from the SDE~\eqref{eq BS SDE},
where $N_\text{steps}=50$. The assets' initial values $s_{t_0}^i$ are  sampled from a lognormal distribution 
\[
X \sim \exp ((\mu-0.5\sigma^2)\tau + \sigma\sqrt{\tau}\xi),
\]
where $\xi\sim N(0,1)$, $\mu=0.08$, $\tau=0.1$.

We follow the evaluation framework to evaluate the model, simulating $N_{\text{MC}} \cdot N_{\text{in}}$ paths by 
simulating~\eqref{eq BS SDE} with constant $S^i_0 = 1$ for $i=1,\ldots,100$. We have the following results:
\begin{enumerate}[i)]
\item Table~\ref{table results empirical variance exchange 100d} provides the empirical variances calculated over $10^6$ generated Monte Carlo samples and their corresponding control variates. 
The variance reduction measure indicates the quality of each control variate method. 
Table~\ref{table results empirical variance exchange 100d}
also provides the cost of training for each method, given by the number of optimiser iterations performed before 
hitting the stopping criteria with $\epsilon = 5 \cdot 10^{-6}$. 
Algorithm~\ref{alg PDE prob repr iterative training} outperforms the other algorithms in terms of variance reduction factor. This is not surprising as Algorithm~\ref{alg PDE prob repr iterative training} explicitly learns the discretisation of the Martingale representation (equation~\eqref{eq lin after ito and eq}) from which the control variate arises. 
\item Table~\ref{table results conf interval exchange 100d} provides the confidence interval for the variance 
of the Monte Carlo estimator, and the Monte Carlo estimator with control variate assuming these 
are calculated on $10^6$ random paths.
\item Figures~\ref{fig var red it eps exchange 100dim} and~\ref{fig it eps exchange 100dim} study the iterative training 
for the BSDE solver. We train the same model four times for different values of $\epsilon$ 
between $0.01$ and $5\times 10^{-6}$ and we study the number of iterations necessary to meet the stopping criteria
defined by $\epsilon$, the variance reduction once the stopping criteria is met, and the relationship between the 
number of iterations and the variance reduction. Note that in this case the variance reduction does not stabilise  for $\epsilon<10^{-5}$. However, the number of training iterations increases exponentially as $\epsilon$ decreases, 
and therefore we also choose $\epsilon = 5\times 10^{-6}$ to avoid building a control that requires a high number of
random paths to be trained.
\end{enumerate}

\begin{table}[H]
\begin{tabular}{|l||c|c|c|c|} 
\hline 
Method & Emp. Var. & Var. Red. Fact. & Train. Paths & Opt. Steps \\ 
\hline \hline
Monte Carlo & $1.97\times 10^{-2}$ & - & - & - \\
\hline 
Algorithm~\ref{alg PDE cond expec}  & $5.94 \times 10^{-3}$  &  $33.16$ & $74 \times 10^6$ & $14\, 900$  \\ 
\hline

Algorithm~\ref{alg PDE prob repr iterative training} & $1.51\times 10^{-4}$ & $130.39$ & $14.145 \times 10^6$ & $2\,829$ \\ 
\hline
Algorithm~\ref{alg empirical risk minimisation} & $5.29 \times 10^{-4}$ & $37.22$ & $97.265\times 10^6$ & $19\,453$  \\ 
\hline 
Algorithm~\ref{alg empirical corr maximisation} &  $1.93\times 10^{-4}$ & $102.05$ & $76.03\times 10^6$  & $15\,206$  \\ 
\hline
\end{tabular} 
\caption{Results on exchange option problem on 100 assets, 
Example~\ref{ex 100d average exchange}. 
Empirical Variance and variance reduction factor and costs in terms 
of paths used for training and optimizer steps.}
\label{table results empirical variance exchange 100d} 
\end{table}

\begin{table}[H]
\centering
\begin{adjustbox}{max width=\textwidth}
\begin{tabular}{@{}|l||c|c|} 
\hline 
Method & Confidence Interval Variance & Confidence Interval Estimator  \\ 
\hline \hline
Monte Carlo & $[1.51\times 10^{-6}, 2.65\times 10^{-6}]$ & $[0.0845,0.0849]$ \\
\hline 
Monte Carlo + antithetic paths & $[8.77\times 10^{-7}, 1.53\times 10^{-6}]$ & $[0.0845,0.0848]$ \\
\hline 

Algorithm~\ref{alg PDE cond expec} & $[3.04 \times 10^{-8}, 2.14 \times 10^{-7}]$ & $[0.0848,0.08493]$\\
\hline
Algorithm~\ref{alg PDE prob repr iterative training} & $[5.32\times 10^{-9}, 1.41\times 10^{-8}]$ & $[0.08485,0.8492]$ \\
\hline
Algorithm~\ref{alg empirical risk minimisation} & $[4.13\times 10^{-9}, 1.09\times 10^{-8}]$ & $[0.08484,0.08490]$ \\
\hline 
Algorithm~\ref{alg empirical corr maximisation} & $[3.80\times 10^{-9}, 1.0\times 10^{-8}]$ & $[0.08487,0.08493]$ \\
\hline
\end{tabular} 
\end{adjustbox}
\caption{Results on exchange option problem on 100 assets,
Example~\ref{ex 100d average exchange}.}
\label{table results conf interval exchange 100d} 
\end{table}  



\begin{figure}[H]
\includegraphics[width=0.5\linewidth]{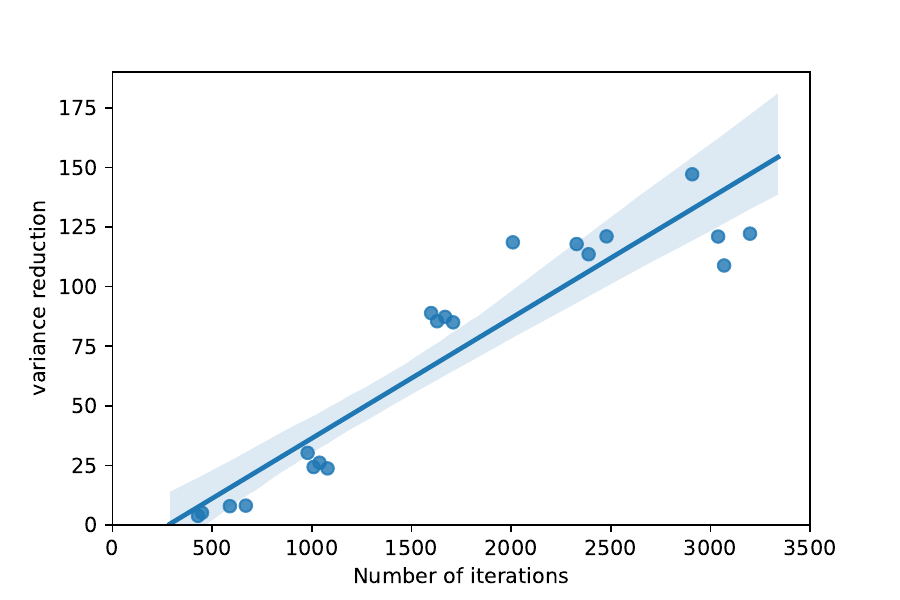}\includegraphics[width=0.5\linewidth]{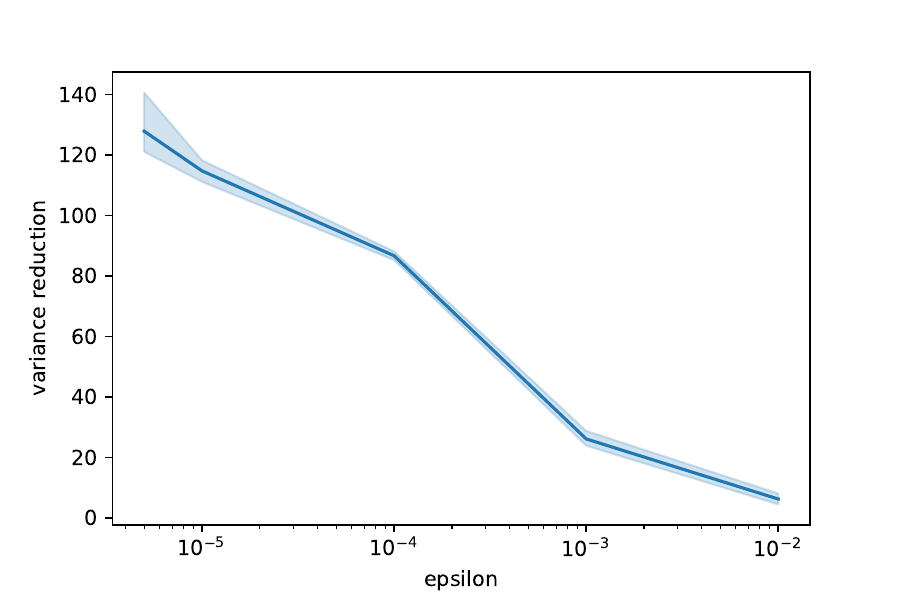}
\caption{Left: Variance reduction in terms of number of optimiser iterations. Right: Variance reduction in terms of epsilon.
Both for Example~\ref{ex 100d average exchange} and
Algorithm~\ref{alg PDE prob repr iterative training}.}
\label{fig var red it eps exchange 100dim}
\end{figure}

\begin{figure}[H]
\includegraphics[width=0.5\linewidth]{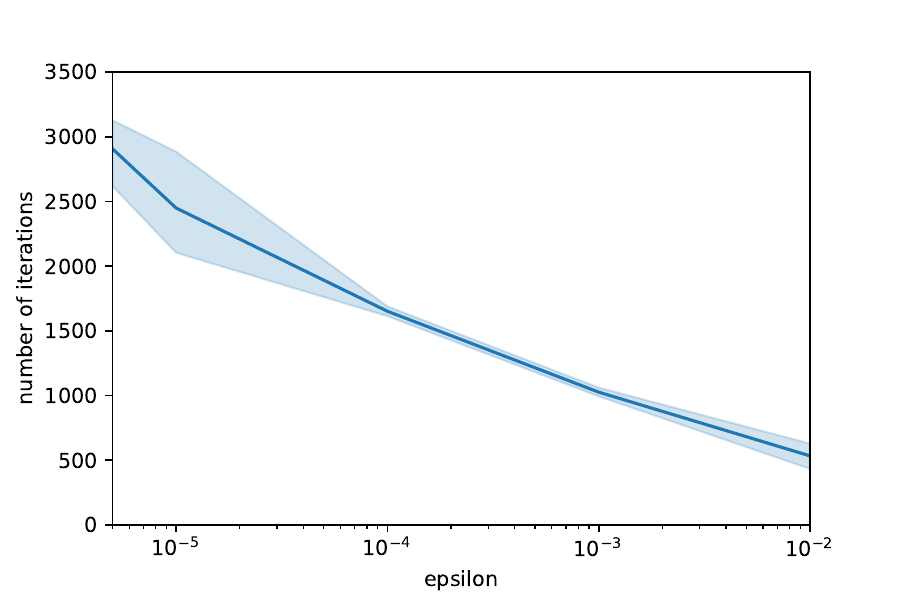}
\caption{Number of optimiser iterations in terms of $\epsilon$ 
for Example~\ref{ex 100d average exchange} and
Algorithm~\ref{alg PDE prob repr iterative training}.
}
\label{fig it eps exchange 100dim}
\end{figure}

\begin{figure}[H]
\includegraphics[width=0.8\linewidth]{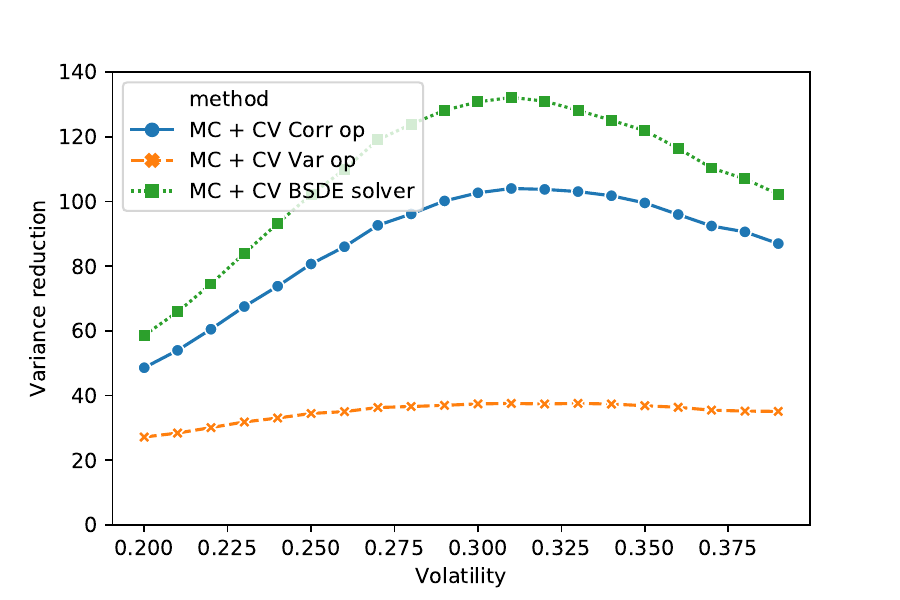}
\caption{Variance reduction with network trained with $\sigma = 0.3$ but applied for $\sigma \in [0.2,0.4]$ for the model of Example~\ref{ex 100d average exchange}.
We see that the variance reduction factor is considerable even in the case when the network is used with ``wrong'' $\sigma$.
It seems that Algorithm~\ref{alg empirical risk minimisation} is not performing well in this case.}
\label{fig var reduction sigmas exchange 100dim}
\end{figure}

\end{example}

\appendix
\section{Martingale Control Variate}\label{sec martingale cv}
Let $(\Omega, \mathcal F, \mathbb P)$ be a probability space 
and consider an $\mathbb R^{d'}$-valued Wiener process $W=(W^j)_{j=1}^{d'} = ((W^j_t)_{t\geq 0})_{j=1}^{d'}$.
We will use $(\mathcal F^W_t)_{t\geq 0}$ to denote the filtration generated by $W$. 
Consider a $D\subseteq \mathbb R^d$-valued, continuous, stochastic process defined for the parameters $\beta\in B\subseteq\mathbb R^p$, $X^{\beta}=(X^{\beta,i})_{i=1}^d = ((X^{\beta,i}_t)_{t\geq 0})_{i=1}^d$ adapted to $(\mathcal F^W_t)_{t\geq 0}$.

Let $g: C([0,T], \mathbb R^d) \to \mathbb R$ be a measurable function. We shall consider path-dependent contingent claims of the form $g((X_s^\beta)_{s\in[0,T]})$. Finally we assume that there is a (stochastic) discount factor given by
\[
D(t_1,t_2;\beta) = e^{-\int_{t_1}^{t_2} c(s,X_s^\beta;\beta)ds}
\]
for an appropriate function $c=c(t,x;\beta)$. We will omit the $\beta$ from the discount factor notation. Let
\[
\Xi_T^\beta := D(t,T)g((X_s^\beta)_{s\in[0,T]})\, .
\]
We now interpret $\mathbb P$ as some risk-neutral measure and so the $\mathbb P$-price of our contingent claim is
\[
V_t^{\beta} = \mathbb E\left[\Xi_T^\beta \big| \mathcal F_t^\beta \right] = \mathbb E\left[D(t,T)g((X_s^\beta)_{s\in[0,T]}) \big| \mathcal F_t^\beta \right]\, .
\]
By assumption $\Xi_T^\beta$ is $\mathcal F_T^W$ measurable and $\mathbb E [|\Xi_T^\beta|^2] <\infty$. Hence, from the Martingale Representation Theorem, see e.g.~\cite[Th. 14.5.1]{cohen2015stochastic}, there exists a unique process $(Z_t^\beta)_{t}$ adapted to $(\mathcal F_t^W)_t$ with $\mathbb E[ \int_0^T |Z_s^\beta|^2 ds ] < \infty$ such that
\begin{equation}\label{eq mrt}
\Xi_T^\beta = \mathbb E[ \Xi_T^\beta \big| \mathcal F^W_0  ] + \int_0^T Z_s^\beta \,dW_s\,.
\end{equation}
The proof of the existence of the process $(Z_t^\beta)_{t}$, is non-constructive. In the setup of the paper, we used the Markovian property of $\Xi_t^\beta$ to approximate $Z_t^\beta$ via the associated linear PDE. In the more general non-Markovian setup, \cite{Cont_2016} provides a numerical method to construct the martingale representation. 

Observe that in our setup, $\mathcal F_0 =\mathcal F_0^W$, $\mathcal F_t^\beta \subseteq \mathcal F_t^W$ for $t\geq 0$.
Hence tower property of the conditional expectation implies that  
\begin{equation} \label{eq mrt 2}
\mathbb E[\Xi_T^\beta \big| \mathcal F_t^\beta ]  = \mathbb E[ \Xi_T^\beta \big| \mathcal F^W_0  ] + \int_0^t Z_s^\beta \, dW_s\,.
\end{equation}
Consequently~\eqref{eq mrt} and~\eqref{eq mrt 2} imply
\[
\mathbb E[\Xi_T^\beta \big| \mathcal F_t^\beta ] = \Xi_T^\beta -  \int_t^T Z_s^\beta \,dW_s\,.
\]
We then observe that 
\[
V_t^\beta = \mathbb E[\Xi_T^\beta \big| \mathcal F_t^\beta ] = \mathbb E\left[\Xi_T^\beta -  \int_t^T Z_s^\beta \, dW_s \bigg| \mathcal F_t^\beta \right]\,. 
\]

If we can generate iid $(W^i)_{i=1}^N$ and $(Z^{\beta,i})_{i=1}^N$ with the
same distributions as $W$ and $Z$ respectively then we can consider
the following Monte-Carlo estimator of $V_t^\beta$: 

\[
\mathcal  V^{\beta, N}_t := \frac1{N}\sum_{i=1}^{N}\left( \Xi_T^{\beta,i} -  \int_t^T Z^{\beta,i}_s \, dW^i_s \right) \,. 
\]

In the companion paper~\cite{sabatevidales2020solving} we provide deep learning algorithms to price 
path-dependent options in the risk neutral measure by solving the corresponding path-dependent PDE, using a combination of Recurrent Neural networks and path signatures to parametrise the process $Z^\beta$.

\section{Martingale Control Variate Deep Solvers}\label{sec appendix martingale control variate deep solvers}
\subsection{Empirical correlation maximisation}
This method is based on the idea that since we are looking for a good control variate 
we should directly train the network to maximise the variance reduction 
between the vanilla Monte-Carlo estimator and the control variates Monte-Carlo estimator
by also trying to optimise $\lambda$.

Recall we denote  $\Xi_T = D(t,T) g((X_s)_{s\in[t,T]})$. We also denote as $M_{t,T}$ as the stochastic integral that arises in the martingale representation of $\Xi_T$. The optimal coefficient $\lambda^{*,\theta}$ that minimises the variance 
$\mathbb{V}\text{ar} [\Xi_T - \lambda M^{\theta}_{t,T}]$ is 
\[
\lambda^{*,\theta} = \frac{ \mathbb{C}\text{ov}[\Xi_T,  M^{\theta}_{t,T}]}{\mathbb{V}\text{ar}[ M^{\theta}_{t,T}]}\,.
\]
Let $\rho^{\Xi_T , M^{\theta}_{t,T} }$ denote the Pearson correlation coefficient between 
$\Xi_T$ and $M^{\theta}_{t,T}$ i.e. 
\[
\rho^{\Xi_T , M^{\theta}_{t,T} } = \frac{\mathbb C\text{ov}(\Xi_T, M^{\theta}_{t,T})}{\sqrt{\mathbb{V}\text{ar}[ \Xi_T]\mathbb{V}\text{ar}[ M^{\theta}_{t,T}]}}\,.
\]
With the optimal $\lambda^*$ we then have that the variance reduction obtained from the control variate is
\[
\frac{\mathbb{V}\text{ar}[ \mathcal V^{\pi,\theta,\lambda^*,N}_{t,T}  ]}{\mathbb{V}\text{ar} [\Xi_T ]} = 1 -\bigg( \rho^{\Xi_T , M^{\theta}_{t,T} } \bigg)^2\,.
\]
See \cite[Ch. 4.1]{glasserman2013monte} for more details. 
Therefore we set the learning task as:
\[
\theta^{*,cor}:=  \argmin_{\theta} \left [1 -\bigg( \rho^{\Xi_T , M^{\theta}_{t,T} } \bigg)^2\ \right]\,.
\]
The implementable version requires the definition of $\mathcal V^{\beta,\pi,\theta,\lambda,N}_{t,T}$ in~\eqref{eq cv mc},
where we set
\[
\begin{split}
	\Xi_T^{\beta,\pi,i} := &  D^{\pi}(t,T))^ig(X^{\beta,\pi,i}_T) \\
	M_{t,T}^{\beta,\pi,i,\theta} := & \sum_{k=1}^{N_\text{steps}-1} (D^{\pi}(t,t_k))^i \mathcal R[\partial_x v]_\theta(t_k,X^{\beta,\pi,i}_{t_k}) \sigma(t_k,X^{\beta,\pi,i}_{t_k})\,(W^i_{t_{k+1}} - W^i_{t_{k}})
\end{split}
\]

The full method is stated as Algorithm~\ref{alg empirical corr maximisation}.

\begin{algorithm}
\caption{Martingale control variates solver: Empirical correlation maximization}
\label{alg empirical corr maximisation}
\begin{algorithmic}
\STATE{Initialisation: $\theta$, $N_{\text{trn}}$}
\FOR{$i:1:N_{\text{trn}}$}
\STATE{ 
generate  samples 
$(x_t^{\beta,\pi,i})_{t\in\pi}$ by using numerical SDE solver on \eqref{eq sde} and sampling from the distribution of $\beta$.
}
\ENDFOR
\STATE{Find $\theta^{*,N_{\text{trn}}}$ using SGD where 
\[
\begin{split}
 \theta^{*,N_{\text{trn}}}:= & \argmin_{\theta} \left[ 1- 
  \left(\frac{\mathbb C \text{ov}^{N_{\text{trn}}}(\Xi_T^{\beta,\pi})}{\sqrt{\mathbb{V}\text{ar}^{N_{\text{trn}}}[ \Xi_T^{\beta,\pi}]\mathbb{V}\text{ar}^{N_{\text{trn}}}[ M^{\beta,\pi,\theta}_{t,T}]}}\right)^2\,\,\right]\,,
\end{split}
\]
where $\mathbb{V}\text{ar}^{N_{\text{trn}}}, \mathbb C \text{ov}^{N_{\text{trn}}}$ denote the empirical variance and covariance. 
 }
\RETURN $\theta^{*,N_{\text{trn}}}$.
\end{algorithmic}
\end{algorithm}

\section{Artificial neural networks}
\label{sec dn}
We fix a locally Lipschitz function $\mathbf a:\mathbb R \to \mathbb R$ and
for $d\in \mathbb N$ define $\mathbf A_d : \mathbb R^d \to \mathbb R^d$ as the function given, for $x=(x_1,\ldots,x_d)$ by 
$\mathbf A_d(x) = (\mathbf a(x_1),\ldots, \mathbf a(x_d))$.
We fix $L\in \mathbb N$ (the number of layers), $l_k \in \mathbb N$, $k=0,1,\ldots L-1$ (the size of input to layer $k$) and $l_L \in \mathbb N$ (the size of the network output). 
A fully connected artificial neural network is then given by $\Phi = ((W_1,B_1), \ldots, (W_L, B_L))$,
where, for $k=1,\ldots,L$, we have real $l_{k-1}\times l_k$ matrices $W_k$ and real $l_k$ dimensional
vectors $B_k$.

The artificial neural network defines a function $\mathcal R_{\Phi} : \mathbb R^{l_0} \to \mathbb R^{l_L}$ given recursively, for $x_0 \in \mathbb R^{l_0}$, by 
\[
\mathcal R_\Phi(x_0) = W_L x_{L-1} + B_L\,, \,\,\,\, 
x_k = \mathbf A_{l_k}(W_k x_{k-1} + B_k)\,,k=1,\ldots, L-1\,.
\]
We can also define the function $\mathcal P$ which counts the number of parameters as 
\[
\mathcal P(\Phi) = \sum_{i=1}^L (l_{k-1}l_k + l_k )\,.
\]
We will call such class of fully connected artificial neural networks $\mathcal{DN}$.
Note that since the activation functions and architecture are fixed the learning task entails
finding the optimal $\Phi \in \mathbb R^{\mathcal P(\Phi)}$.

\section{Additional numerical results}
\begin{example}[Low dimensional basket option]
\label{ex basket}
We consider the basket options problem of pricing, using the example from \cite[Sec 4.2.3]{belomestny2017variance}.
The payoff function is
\[
g(S):= \max\left(0,\sum_{i=1}^d S^i - K \right)\,.
\]
We first consider the basket options problem on two assets, with 
$d=2$, $S^i_0 = 70$, $r=50\%$, $\sigma^i = 100\%$, $\Sigma^{ii} = 1$, $\Sigma^{ij} = 0$ for $i\neq j$, and constant strike $K = \sum_{i=1}^{d} S^i_0$. 
In line with the example from \cite[Sec 4.2.3]{belomestny2017variance} for comparison purposes we organise the experiment as follows. 
The control variates on $20\,000$ batches of $5\,000$ samples each of  $(s_{t_n}^i)_{n=0}^{N_\text{steps}}$ by simulating the SDE~\ref{eq BS SDE},
where $N_\text{steps}=50$. The assets' initial values $s_{t_0}$ are always constant $S_{t_0}^i = 0.7$. 
We follow the evaluation framework to evaluate the model, simulating $N_{\text{MC}} \cdot N_{\text{in}}$ paths by 
simulating~\ref{eq BS SDE} with constant $S^i_0 = 0.7$ for $i=1,\ldots,100$. 
We have the following results:
\begin{enumerate}[i)]
\item Table~\ref{table results basket constant 2d} provides the empirical variances calculated over $10^6$ generated Monte Carlo samples and their corresponding control variates. 
The variance reduction measure indicates the quality of each control variate method. 
Table~\ref{table results basket constant 2d}
also provides the cost of training for each method, given by the number of optimiser iterations performed before 
hitting the stopping criteria, defined defined before with $\epsilon = 5\times 10^{-6}$. 
\item Table~\ref{table results conf interval basket constant 2d} provides the confidence interval for the variance 
of the Monte Carlo estimator, and the Monte Carlo estimator with control variate assuming these 
are calculated on $10^6$ random paths.
\item Figures~\ref{fig var red it eps basket 2dim} and~\ref{fig it eps basket 2dim} study the iterative training 
for the BSDE solver. 
We train the same model four times for different values of $\epsilon$ 
between $0.01$ and $5\times 10^{-6}$ and we study the number of iterations necessary to meet the stopping criteria
defined by $\epsilon$, the variance reduction once the stopping criteria is met, and the relationship between the 
number of iterations and the variance reduction. 
Note that the variance reduction stabilises  for $\epsilon<10^{-5}$. Furthermore, the number of training iterations increases exponentially as $\epsilon$ decreases.
We choose $\epsilon = 5\times 10^{-6}$.
\end{enumerate}

We note that in the example from \cite[Sec 4.2.3]{belomestny2017variance}, the control variate is trained with $S_0=0.7$ fixed. Using this setting, Algorithm~\ref{alg PDE cond expec} cannot be used to approximate the control variate in~\eqref{eq cv mc}: since the network at $t=0$, $\mathcal R[v]_{\eta_0}$, is trained only at $S_0=0.7$, then automatic differentiation to approximate $\partial_x \mathcal R[v]_{\eta_0}(0.7)$ will yield a bad approximation of $\partial_x v(0.7)$; indeed, during training $\mathcal R[v]_{\eta_0}$ is unable to capture how $v$ changes around $S_0$ at $t=0$. For this reason, Algorithm~\ref{alg PDE cond expec} is not included in the following results.

\begin{table}[h!]
\centering
\begin{tabular}{@{}|l||c|c|c|c|} 
\hline 
Method & Emp. Var. & Var. Red. Fact. & Train. Paths & Optimizer steps \\ 
\hline \hline
Monte Carlo & $1.39$ & - & - & - \\
\hline 
Algorithm~\ref{alg PDE prob repr iterative training} & $1.13\times 10^{-3}$ & $1219$ & $ 8 \times 10^7$ & $16\,129$   \\ 
\hline 
Algorithm~\ref{alg empirical risk minimisation} & $1.13 \times 10^{-3}$ & $1228$ & $3\times 10^7$ & $6\,601$  \\ 
\hline 
Algorithm~\ref{alg empirical corr maximisation} &  $1.29\times 10^{-3}$ & $1076$ & $4\times 10^7$ & $8\,035$ \\ 
\hline
\end{tabular} 
\caption{Results on basket options problem on two assets, Example~\ref{ex basket}.
Models trained with $S_0$ fixed, non-random. 
Empirical Variance and variance reduction factor are presented.}
\label{table results basket constant 2d} 
\end{table}

\begin{table}[h!]
\centering
\begin{tabular}{@{}|l||c|c|} 
\hline 
Method & Confidence Interval Variance & Confidence Interval Estimator  \\ 
\hline \hline
Monte Carlo & $[4.49\times 10^{-5}, 1.19\times 10^{-4}]$ & $[0.665,0.671]$ \\
\hline 
Monte Carlo + antithetic paths & $[1.43\times 10^{-5}, 2.51\times 10^{-5}]$ & $[0.667,0.670]$ \\
\hline 

Algorithm~\ref{alg PDE prob repr iterative training} & $[2.1329\times 10^{-8}, 5.6610\times 10^{-8}]$ & $[0.6696,0.6697]$ \\
\hline 
Algorithm~\ref{alg empirical risk minimisation} & $[1.687\times 10^{-8}, 4.47\times 10^{-7}]$ & $[0.6695,0.6697]$ \\
\hline 
Algorithm~\ref{alg empirical corr maximisation} & $[1.746\times 10^{-8}, 4.63\times 10^{-8}]$ & $[0.6695,0.6697]$ \\
\hline
\end{tabular} 
\caption{Results on basket options problem on two assets, Example~\ref{ex basket}. 
Models trained with $S_0$ fixed, non-random. }
\label{table results conf interval basket constant 2d} 
\end{table}  



\begin{figure}[H]
\includegraphics[width=0.5\linewidth]{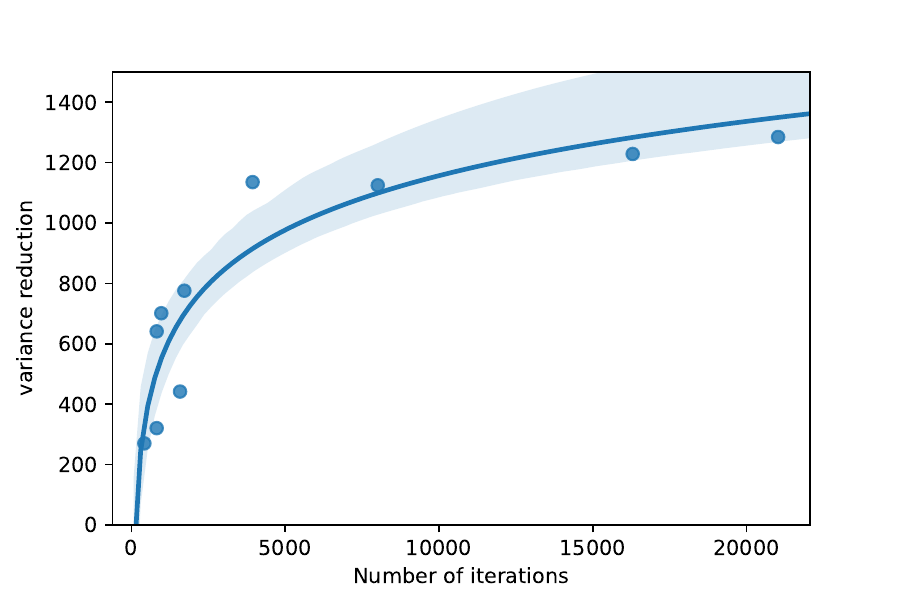}\includegraphics[width=0.5\linewidth]{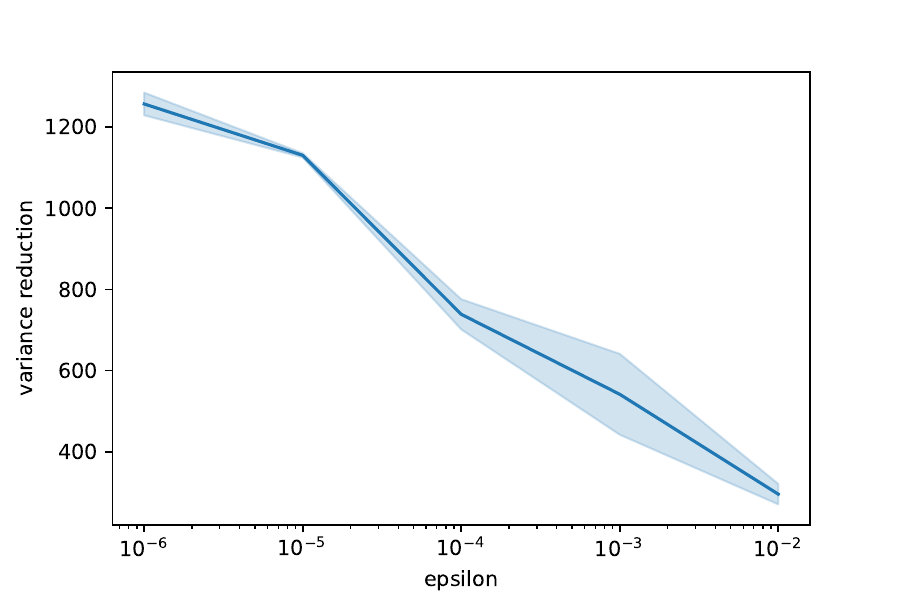}
\caption{Left: Variance reduction in terms of number of optimiser iterations. Right: Variance reduction in terms of epsilon.
Both refer to Algorithm~\ref{alg PDE prob repr iterative training} used
in Example~\ref{ex basket}.}
\label{fig var red it eps basket 2dim}
\end{figure}

\begin{figure}[H]
\includegraphics[width=0.5\linewidth]{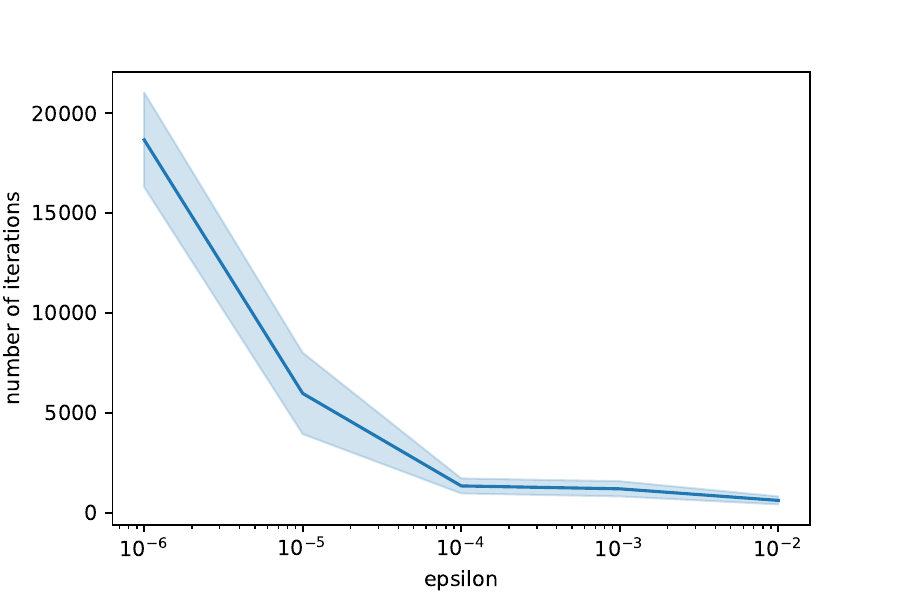}
\caption{Number of optimiser iterations in terms of $\epsilon$
for Algorithm~\ref{alg PDE prob repr iterative training} used
in Example~\ref{ex basket}.}
\label{fig it eps basket 2dim}
\end{figure}

\end{example}

\begin{example}[basket option with random sigma]
\label{ex basket random sigma}
In this example, as in Example~\ref{ex 2d exchange price approx vs CV}, we aim to show how our approach  - where we build a control variate 
by approximating the process $(Z_{t_k})_{k=0,\ldots,N_{\text{steps}}}$ - is more robust compared to directly approximating the price by a certain function 
in a high-dimensional setting. 

We use the methodology proposed in \cite{Horvath2019DeepVolatility}, where the authors present a deep learning-based calibration 
method proposing a two-steps approach: first the authors learn the model that approximates the pricing map using a artificial neural network in which the inputs 
are the parameters of the volatility model. Second the authors calibrate the learned model using available data by means of different optimisation methods. 

For a fair comparison between our deep learning based control variate approach vs. the method proposed in \cite{Horvath2019DeepVolatility},
we make the following remarks:
\begin{enumerate}[i)]
\item We will only use the the first step detailed in \cite{Horvath2019DeepVolatility} where the input to the model that approximates the pricing map are 
the volatility model's parameters: $\sigma \in \mathbb R^d, r$, and the initial price is considered constant for training purposes. 
We run the experiment for $d=5$.
\item In \cite{Horvath2019DeepVolatility} the authors  build a training set, and then perform gradient descent-based optimisation 
on the training set for a number of epochs. This is somewhat a limiting factor in the current setting where one can have as much data as they want 
since it is generated from some given distributions. In line with our experiments, instead of building a training set, in each optimisation step we sample
a batch from the given distributions. 
\item In \cite{Horvath2019DeepVolatility}, the price mapping function is learned for a grid of combinations of maturities and strikes. In this experiment, we 
reduce the grid to just one point considering $T=0.5$, $K=\sum_i S_0^i$, where $S_0^i=0.7\, \, \forall i$.
\item We will use Algorithm~\ref{alg PDE prob repr iterative training} to build the control variate with the difference that now $\sigma \in \mathbb R^d, r\in\mathbb R$ will be passed as
input to the each network at each time step $\mathcal R[v]_{\eta_k}, \mathcal R[\partial_x v]_{\theta_k}$. 
\end{enumerate}

The experiment is organised as follows:

\begin{enumerate}[i)]
\item We train the network proposed in \cite{Horvath2019DeepVolatility} approximating the price using Black-Scholes model
and Basket options payoff. In each optimisation iteration a batch of size 
$1\, 000$, where the volatility model's parameters are sampled using $\sigma \sim \mathcal{U}(0.9,1.1)$ and $r \sim \mathcal{U}(0.4,0.6)$. 
We keep a test set of size 
150, $\mathcal S = \{\left[(\sigma^i, r^i); p(\sigma^i,r^i)\right],i=1,\ldots,150\}$ where $p(\sigma^i,r^i)$ denotes the price and
is generated using $50\, 000$ Monte Carlo samples. 
\item We use Algorithm~\ref{alg PDE prob repr iterative training}  to build the control variate, where $\sigma$ and $r$  are
sampled as above and are included as inputs to the network. We denote the trained model by 
$\mathcal R[\partial_x v]_{\theta_k}$ where $k=1,\ldots,N_{\text{steps}}$. In contrast with Algorithm~\ref{alg PDE prob repr iterative training}.
\end{enumerate}

We present the following results:
\begin{enumerate}[i)]
\item Figure~\ref{fig hist MSE test set} displays the histogram of the squared error of the approximation of the PDE solution $\mathcal R[v]_\eta$ for each instance in $\mathcal S$. 
In this sample, it spans from almost $10^{-8}$ to $10^-3$, i.e. for almost five orders of magnitude. 
\item We build the control variate for that instance in the test set for which $\mathcal R[v]_\eta$
generalises the worst. For those particular $\sigma, r$, Table~\ref{table results empirical variance basket constant 5d} provides its variance reduction factor. 
\end{enumerate}


\begin{table}[H]
\begin{tabular}{@{}|c||c|c|c|c|} 
\hline 
Method & Emp. Var. & Var. Red. Fact.  \\ 
\hline  \hline
Monte Carlo & $1.29$ & - \\
\hline 
Algorithm~\ref{alg PDE prob repr iterative training} & $0.035$ & $37$  \\ 
\hline
\end{tabular} 
\caption{Results on basket options problem on 5 assets, 
Model trained with non-random $S_0$, and random $\sigma, r$.}
\label{table results empirical variance basket constant 5d} 
\end{table}

\end{example}

\begin{example}[High dimensional basket option]
\label{ex basket 100d}
We also consider the basket options problem on $d=100$ assets
but otherwise identical to the setting of Example~\ref{ex basket}.
We compare our results against
 the same experiment in \cite[Sec 4.2.3, Table 6 and Table 7]{belomestny2017variance}. 

\begin{table}[H]
\begin{tabular}{@{}|c||c|c|c|c|} 
\hline 
Method & Emp. Var. & Var. Red. Fact. & Train. Paths & Opt. Steps \\ 
\hline  \hline
Monte Carlo & $79.83$ & - & - & -\\
\hline 
Algorithm~\ref{alg PDE prob repr iterative training} & $4.72\times 10^{-4}$ & $168\,952$ & $24\times 10^7$  & $47369$ \\ 
\hline
Algorithm~\ref{alg empirical risk minimisation} & $1.79 \times 10^{-4}$ & $349\,525$ & $37\times 10^6$ & $7383$ \\ 
\hline 
Algorithm~\ref{alg empirical corr maximisation} &  $1.54\times 10^{-4}$ & $517\,201$ & $35\times 10^6$ & $7097$ \\ 
\hline
Method $\zeta_{\text{a}}^1$ in~\cite{belomestny2017variance} & $8.67\times 10^{-1}$  & $97$ & - & - \\
\hline
Method $\zeta_{\text{a}}^2$ in~\cite{belomestny2017variance} & $4.7\times 10^{-3}$ & $17\,876$ & - & - \\ 
\hline
\end{tabular} 
\caption{Results on basket options problem on 100 assets, 
Example~\ref{ex basket 100d}. 
Models trained with non-random $S_0$ so that the results can be directly
compared to~\cite{belomestny2017variance}.}
\label{table results empirical variance basket constant 100d} 
\end{table}  
Table~\ref{table results empirical variance basket constant 100d} shows a significant improvement of the variance reduction factor (10x and 100x better) of all our Algorithms than the methods proposed in~\cite{belomestny2017variance} and applied in the same example.

\begin{table}[H]
\centering
\begin{adjustbox}{max width=\textwidth}
\begin{tabular}{@{}|l||c|c|} 
\hline 
Method & Confidence Interval Variance & Confidence Interval Estimator  \\ 
\hline \hline
Monte Carlo & $[8.57\times 10^{-4}, 2.27\times 10^{-3}]$ & $[27.351,27.380]$ \\
\hline 
Monte Carlo + antithetic paths & $[5.34\times 10^{-4}, 9.35\times 10^{-4}]$ & $[27.354,27.371]$ \\
\hline 

Algorithm~\ref{alg PDE prob repr iterative training} & $[7.001\times 10^{-9}, 1.8583\times 10^{-8}]$ & $[27.3692, 27.3693]$ \\
\hline 
Algorithm~\ref{alg empirical risk minimisation} & $[2.41\times 10^{-9}, 6.39\times 10^{-9}]$ & $[27.36922, 27.36928]$ \\
\hline 
Algorithm~\ref{alg empirical corr maximisation} & $[4.1672\times 10^{-9}, 1.1060\times 10^{-8}]$ & $[27.36922, 27.36928]$ \\
\hline
\end{tabular} 
\end{adjustbox}
\caption{Results on basket options problem on 100 assets, 
Example~\ref{ex basket 100d}. 
Models trained with non-random $S_0$.}
\label{table results conf interval basket constant 100d} 
\end{table}

\begin{figure}[H]
\includegraphics[width=0.5\linewidth]{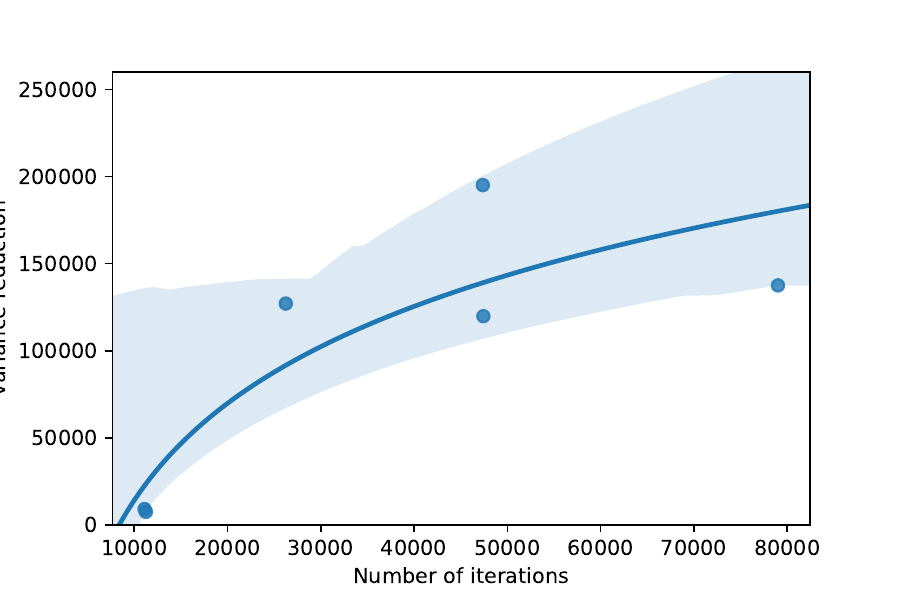}\includegraphics[width=0.5\linewidth]{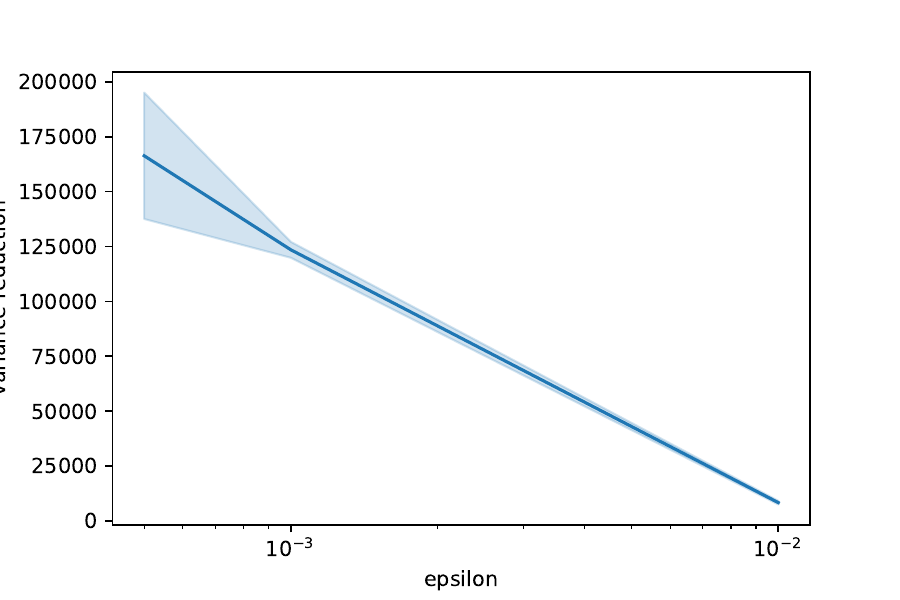}
\caption{Left: Variance reduction in terms of number of optimiser iterations. Right: Variance reduction in terms of epsilon. 
Both are for Example~\ref{ex basket 100d} and Algorithm~\ref{alg PDE prob repr iterative training}.}
\label{fig var red it eps basket 100dim}
\end{figure}


\end{example}

\subsection{Empirical network diagnostics}
\label{sec experiments empirical net diag}
\begin{figure}[H]
\begin{adjustbox}{max width=\textwidth}
\includegraphics[width=.5\linewidth]{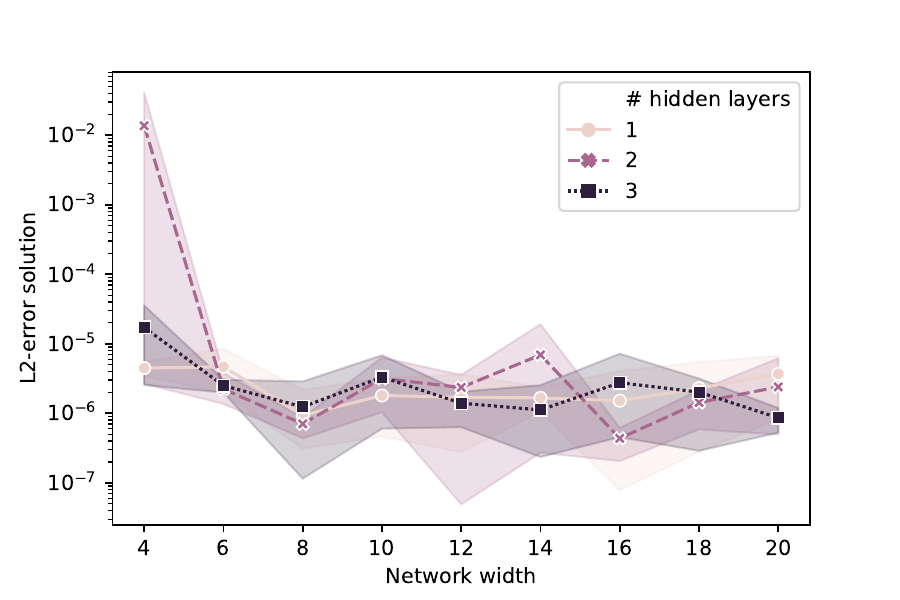}\includegraphics[width=.5\linewidth]{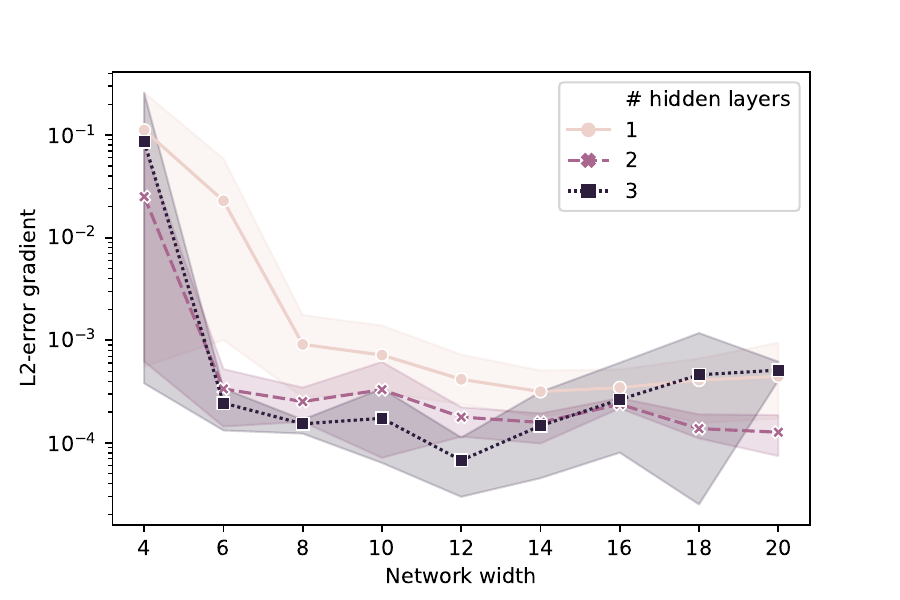}
\end{adjustbox}
\caption{Average error of PDE solution approximation and its gradient and 95\% confidence interval of different combination of \# of layers and net width. Left: error model. Right: Error grad model}
\label{fig error model error grad}
\end{figure}

In this subsection we consider the exchange options problem on two assets from Example~\ref{ex marg}, where the time horizon is one day.
We consider different network architectures for the BSDE method described by 
Algorithm~\ref{alg PDE prob repr iterative training} in order to understand their impact on the final result and their ability to approximate 
the solution of the PDE and its gradient. 
We choose this problem given the existence of an explicit solution that can be used as a benchmark. 
The experiment is organised as follows:
\begin{enumerate}[i)]
\item Let $L-2$ be the number of hidden layers of $\mathcal R[\partial_x v]_{\theta_{t_0}} \in \mathcal{DN}$ and $\mathcal R[v]_{\theta_{t_0}} \in \mathcal{DN}$.
Let $l_k$ be the number of neurones per hidden layer $k$.
\item We train four times all the possible combinations for $L-2 \in \{1,2,3\}$ and for $l_k \in \{2,4,6,\ldots,20\}$ using $\epsilon=5\times 10^{-6}$ for the stopping criteria. 
The assets' initial values $s_{t_0}^i$ are sampled from a lognormal distribution 
\[
X \sim \exp ((\mu-0.5\sigma^2)\tau + \sigma\sqrt{\tau}\xi),
\]
where $\xi\sim N(0,1), \mu=0.08, \tau=0.1$.
\item We approximate the $L^2$-error of $\mathcal R[v]_{\theta_{t_0}}(x)$ and $\mathcal R[v]_{\theta_{t_0}}(x)$ with respect to the exact solution given by Margrabe's formula and its gradient.
\end{enumerate}
Figure~\ref{fig error model error grad} displays the average of the $L^2$-errors and its confidence interval.  
We can conclude that for this particular problem,
the accuracy of $\mathcal R[v]_{\theta_{t_0}}(x)$ does not strongly depend on the number of layers, and that there is no improvement beyond 8 nodes per hidden layer.
The training (its inputs and the gradient descent algorithm together with the stopping criteria) becomes the limiting factor.
The accuracy of $\mathcal R[v]_{\theta_{t_0}}(x)$ is clearly better with two or three
hidden layers than with just one. 
Moreover it seems that there is benefit in taking as many as 10 nodes per hidden
layer.

\section*{Acknowledgements}
This work was supported by the Alan Turing Institute under EPSRC grant no. EP/N510129/1.
              
\raggedright

\bibliographystyle{abbrv}
\bibliography{deep_pde,nested}

\end{document}